\documentstyle[aps,prl]{revtex}
\title{Photon emission in Pb+Pb collisions at SpS and LHC}
\author{D.Yu. Peressounko and Yu.E. Pokrovsky}
\address{RRC "Kurchatov Institute" 123182, Moscow, Kurchatov sq., 1. \\
e-mail: peres@pretty.mbslab.kiae.ru}

\begin{document}
\maketitle

\begin{abstract}

Yield of direct photons in Pb+Pb collisions at SpS and LHC energy is
evaluated with emphasis on estimate of possible uncertainty. Possibility of
experimental observation of direct photons at LHC is discussed. Predictions
of several models at SpS energy are compared with experimental data.

\end{abstract}

\vspace{1cm}

Photon emission is one of the most actively used observables both in present
and developed heavy ion experiments. The reason is that photons, in contrast
to hadronic probes, have extremely large free path length in the hot matter
and escape from it without rescattering. This makes them a unique probe of
the initial stage of collision, and the inner part of the created hot matter.
In particular, direct photons are considered as one of the most promising
signatures of quark-gluon plasma.

As well known, photons, emitted in the heavy ion collision, can be divided
onto direct and decay ones. The direct photons are emitted as a result of
rescatterings of charged particles in the hot matter. The decay photons are
the photons, originated due to decays of final hadrons, mainly $\pi^0$ and
$\eta$ mesons. In addition to this, one can subdivide direct photons on
prompt and thermal. Prompt photons are emitted in the very beginning of the
collision in scatterings of charged constituents of target and projectile.
They have approximately the same energy distribution as emitting particles --
power one -- and thus contribute mainly into the hard part of the spectrum.
Thermal photons are emitted on the later stages of collision, when local
thermodynamic equilibrium is almost reached. They have approximately
exponential spectrum. Decay photons can be subdivided on photons, originated
from decay of hadrons, emitted from QGP surface \cite{we} and from hadronic
gas, and therefore bring out information about two different phases of the
hadron matter.

Despite a large number of papers devoted to photon emission in heavy ion
collisions \cite{Shuryak0}-\cite{WA98-UrQMD}, it is not quite clear, how
predictions of various models differ from each other, how strong is
sensitivity of these predictions to the variations of model parameters,
approximations, and assumptions of the models, and how reasonable are
attempts to find photon signatures of QGP. These points are not discussed in
the literature detailed enough to estimate simultaneously all important
contributions to photon spectra in wide range of photon momentum. In this
paper we made an attempt to compare predictions of various models in the
common basis and estimate sensitivity of the predictions to the variations of
the model parameters. Thus, basing on SpS data, we estimate possible
uncertainties of predictions of direct and decay photon yield in Pb+Pb
collisions at LHC. Having in mind present and developing experiments we
restrict ourselves by $Pb$ nuclei.

The plan of this paper is following. In the first part we consider yield of
prompt photons and estimate uncertainty in their evaluations. In the second
part we consider evaluation of emission rates of thermal photons from QGP and
hadronic gas with and without chemical equilibrium. Having emission rates we
evaluate yield of thermal photons accounting dynamics of the system within
various hydrodynamic models and explore sensitivity to the model
approximations. Then we compare predictions of cascade and hydrodynamic
models at SpS energies. In the third part we consider decay photons and
evaluate ratio Direct/Decay photons for various hydrodynamic and cascade
models. Our results are summarized in the conclusion.

\section{Prompt photons.}

Prompt photons are emitted in first interactions of charged constituents of
target and projectile. Prompt photons reflect dynamics of the very beginning
of collision, and can be used to study structure functions of colliding
nuclei. To evaluate yield of prompt photons one should convolute momentum
distributions of incident partons (structure functions of colliding nuclei)
$F$ with cross sections of elementary collisions $qg\to \gamma q$, $q\bar q\to
\gamma g$, bremsstrahlung etc.

\begin{equation}
\label{prompt-form}
E_\gamma\frac{d\sigma}{d^3p_\gamma}=\sum \limits _{a,b}\int\frac{dx_a\,
dx_b}{x_a x_b} F(a,A;x_a)\,
F(b,B;x_b)\,E_\gamma\frac{d\sigma_{ab\to c\gamma}}{d^3p_\gamma}
\end{equation}

Due to the comparatively high momentum transfer, these cross-sections can be
evaluated within perturbative QCD with reasonable accuracy. The main
uncertainty in the prompt photon yield comes from structure functions:
emission of photon even with as high energy as $E_\gamma\sim 5\, GeV$ in
Pb+Pb collision at $\sqrt{s}=6300\, A\cdot GeV$ is determined by structure
functions in the low $x$ region ($x\sim E_\gamma/\sqrt{s}\sim 0.001$), where
they are not known well yet. In addition, one should take into account
modifications of structure functions in nuclei, what further increase
uncertainty in the yield of prompt photons.

\begin{figure}[th]
\vspace*{9.cm}
\includegraphics{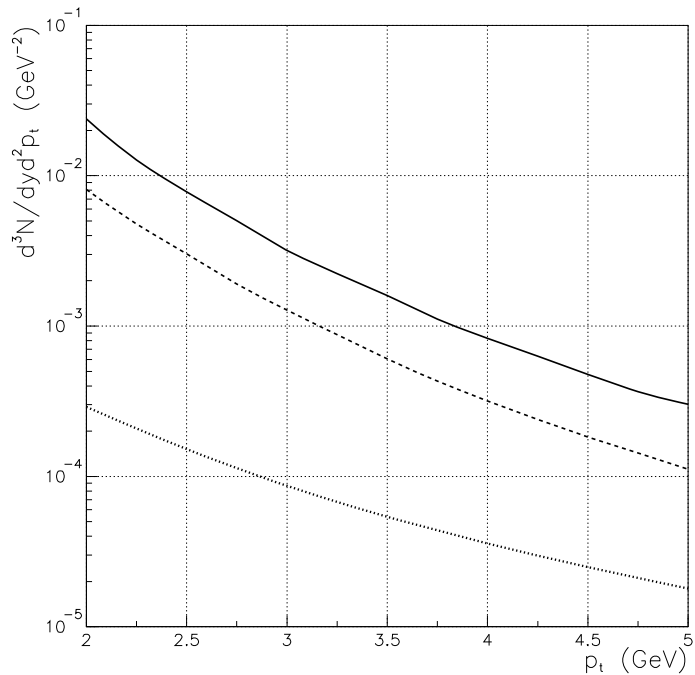}
\caption{Yield of prompt photons in $Pb+Pb$ collision at LHC energy,
         evaluated in \protect\cite{prompt-old,prompt-shad} with structure
         functions of Duke and Owens, set I -- dotted line, with Gluck, Reya
         and Vogt parametrization of structure functions without accounting
         of shadowing (solid line) and with shadowing (dashed  line).
        }
\label{prompt}
\end{figure}

This uncertainty is demonstrated on the fig.\ \ref{prompt}, where
several evaluations with different structure functions are shown. Dotted line
corresponds to the predictions of Alam et al. \cite{prompt-old}. They used
Duke and Owens, set I \cite{Duke} parametrization of structure functions
without accounting any modifications in nuclei. However, since this paper a
new data on the low $x$ behavior of structure functions was obtained, and in
the paper \cite{prompt-shad}, the yield of direct photons was evaluated with
new parametrization (Gluck, Reya and Vogt \cite{Gluck}) of structure
functions. This prediction is shown by solid line. As one can see, various
parametrizations results in difference more then order of magnitude. In
addition to this, structure functions in nuclei significantly differ from
ones of single nucleon: depending on the value of $x$ one finds shadowing,
antishadowing, EMC effect, Fermi motion etc. As far as we are interesting in
low $x$ region, the shadowing is the most important for us. In the paper
\cite{prompt-shad} Gluck-Reya-Vogt structure functions was modified to
include this effect. This result is shown by dashed line.

So that, using of different parametrizations of structure functions
results in difference in the yield of prompt photons up to order of magnitude,
while taking into account modifications of structure functions in nuclei
varies it approximately in three times. On this background one can neglect
uncertainties arising from higher order QCD corrections, which (estimated,
e.g., from $pp$ data) do not exceed a factor 2.

\section{Thermal photons}

Evaluating yield of thermal photons, one assumes, that they are emitted from
a medium at local thermal equilibrium. So that one can average individual
collisions, producing photons, over time and volume and evaluate average
number of photons, emitted from unit volume per unit time -- emission rate
$R$. In the case of perfect thermal equilibrium emission, the rate depends
only on the temperature of the medium, however, in the absence of the
chemical equilibrium it depends also on fugacities of quarks and gluons in
QGP and hadrons in hadronic gas. When emission rates from QGP and hadronic
gas are evaluated, one integrates these rates over space-time volume,
occupied by the system, and finds total yield of thermal photons.

%Local thermal equilibrium is closely related to applicability of hydrodynamic
%description of evolution: local thermal equilibrium suppose numerous
%collisions and thus small free path length -- condition of applicability of
%hydrodynamic description. So, it is more consistent to use hydrodynamics
%rather then e.g. cascade model for description of evolution in evaluating of
%the yield of thermal photons.

%======================================================================

\subsection{Emission rate from Quark-Gluon Plasma}

Thermal photons are emitted from QGP due to reactions: $gq\to\gamma q$
(Compton scattering), $q\bar q\to g \gamma$ (annihilation), $qq\to\gamma qq$
(bremsstrahlung) and other processes with higher order in $\alpha=1/137$ and
$\alpha_s\sim 0.4$ -- the strong coupling constant. To begin with, let us
consider the simplest two-particle reactions:  Compton scattering and
annihilation. In this case, emission rate is described by the following
formula, practically the same, as (\ref{prompt-form}), but with thermal
distributions instead of structure functions, Pauli blocking and Bose
enhancement for final particles:

\begin{eqnarray}
E\frac{dR}{d^4xd^3 p} =\sum\int\prod\limits_{i=1}^{3}
\frac{d^3p_i}{2E_i(2\pi)^3}
f_1(p_1)f_2(p_2)[1\pm f_3(p_3)] \frac{1}{2(2\pi)^3}
|M|^2 (2\pi)^4 \delta^4(p_1+p_2-p_3-p),
\end{eqnarray}
where sum is taken over all possible (two-particle) reactions, $f(p)$ --
thermal distribution functions, $|M|^2$ -- squared matrix element of the
corresponding reaction.

Using $|M|$ evaluated at lowest order both in $\alpha$ and $\alpha_s$, and
taking thermal distributions, one obtains expression for emission rate
\cite{rate-simp}:

\begin{equation}
E\frac{dN}{d^4xd^3 p}=\sum\limits_{u,d,...}e^2_q \frac {\alpha
\alpha_s}{2\pi^2}T^2e^{-E/T} \left [\ln \frac{ET}{m^2} + const \right ].
\end{equation}

\begin{figure}[ht]
\vspace*{9.cm}
\includegraphics{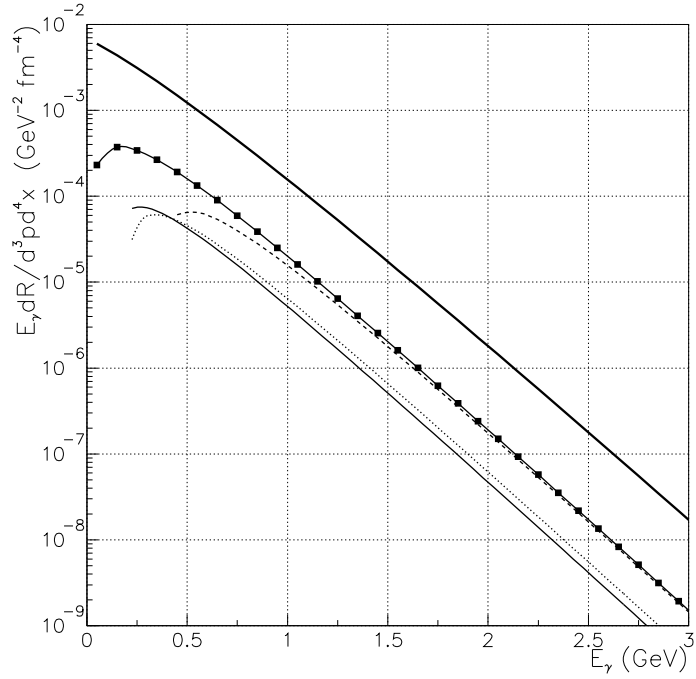}
\caption{Emission rates of thermal photons from QGP at $T=200\, MeV$. Thick
         solid line -- emission rate with accounting of bremsstrahlung and
         annihilation with rescattering on third particle. Solid line with
         rectangles and dashed line correspond to emission rates from Compton
         scattering and annihilation with Bose-Fermi and Boltzmann thermal
         distributions correspondingly. Dotted line and thin solid line --
         chemically nonequilibrated QGP with $\lambda_q=0.5,\, \lambda_g=1$
         and $\lambda_q=0.5,\, \lambda_g=0.5$ correspondingly, only Compton
         scattering and annihilation are taken into account.
         }
\label{rate-QGP-fig}
\end{figure}

This expression diverges for massless quarks $m\to 0$. This is reflection of
the well known Coulomb divergency. To screen it one has to take into account
medium effects. This was done in \cite{rate-htl} using Braaten-Pisarski
technique of hard thermal loops. As a result of this calculations current
mass of quarks $m$ was changed to screening mass $\bar m^2 =g^2 T^2/2.912$
($\alpha_s=g^2/4\pi$) what results in the rate:

\begin{equation}
\label{kapli}
E\frac{dN}{d^4xd^3 p}=\sum\limits_{u,d,...}e^2_q \frac {\alpha \alpha_s}{2\pi^2}T^2e^{-E/T}
\ln \left (\frac{2.912E}{g^2T}\right ).
\end{equation}
Emission rate, evaluated at $T=0.2\, GeV$, $\alpha_s=0.4$ for plasma,
consisting of two flavor quarks and gluons is shown on the fig.\
\ref{rate-QGP-fig} by dashed line. To perform these calculations
analytically, Boltzmann distributions was used instead of Fermi and Bose
ones.  In case of Bose and Fermi distributions one can perform evaluations
numerically. It was shown in \cite{rate-htl} that one can fit this result by
function (\ref{kapli}) with argument of $\ln$ changed to $2.912E/g^2T+1$.
Corresponding curve is shown on the fig.\ \ref{rate-QGP-fig} by solid line,
marked by rectangles.

Let us now return to the more complicated processes -- bremsstrahlung and
annihilation with scattering on third quark. These processes arises in the
next order of $\alpha_s$, compared to Compton scattering and annihilation. It
is well known, that in vacuum the bremsstrahlung contribute mainly into the
soft part of spectrum. So one usually neglects its contribution into thermal
photon emission. However, recently it was shown \cite{rate-twolp}, that in
thermal QCD these processes appear to be important and even dominate over
compton scattering and annihilation even in hard part of spectrum. In this
paper in framework of hard thermal loop technique an expression was obtained
for emission rate of hard ($E_\gamma > T$) photons due to bremsstrahlung:

\begin{equation}
E\frac{dN}{d^4xd^3 p}=\sum\limits_{u,d,...}e^2_q \frac {8}{\pi^5}
\alpha \alpha_s T^2e^{-E/T} (J_T-J_L) \ln 2
\label{twoloop-brems}
\end{equation}

and due to annihilation with scattering on third quark:

\begin{equation}
E\frac{dN}{d^4xd^3 p}=\sum\limits_{u,d,...}e^2_q \frac {8}{3 \pi^5}
\alpha \alpha_s E\, T\,e^{-E/T} (J_T-J_L),
\label{twoloop-ann}
\end{equation}
where, for two quark flavors constants $J_T\approx 4.45$ and $J_L\approx
-4.26$.  It is interesting, that this contributions have the same order in
$\alpha_s$ as two-particle reaction. This takes place because of the presence
of collinear divergency in matrix element of reaction, which is screened by
quark mass.  This results in enhancement factor $\sim 1/m_q^2\sim 1/g^2$.
Sum of these two contributions into thermal photon emission rate is shown on
fig.  \ref{rate-QGP-fig} by thick solid line. As one can see, taking into
account of these processes increases significantly (up to order of magnitude)
emission rate.

If QGP is formed in the nucleus-nucleus collision, then the number of quarks
and gluons in it may be much lower, than in the plasma at perfect chemical
equilibrium. This can strongly affect on the emission rate
\cite{rate-chem1,rate-chem2,rate-mu}. The simplest way to account
undersaturation of quarks in the QGP is just to put coefficient $\lambda_q$
(fugacity of quarks) as a coefficient in the emission rate. More advanced
possibility is in addition to this to evaluate screening masses, depending on
quark and gluon fugacities. However, in the case of absence of chemical
equilibrium one can not be sure that there will be perfect screening of the
column divergency as in the equilibrated case. In the paper \cite{rate-chem2}
it was shown, that even in the case of absence of chemical equilibrium
divergencies cancels each other in the emission rate. In particular case of
distributions of quarks and gluons:

\begin{equation}
f_{g}(p)=\left \{
\begin{array}{ll}
\lambda_g n_B(|p_0|), & p_0>0, \\
-[1+\lambda_gn_B(|p_0|)], & p_0<0
\end{array}
\right .
\qquad
f_{q}(p)=\left \{
\begin{array}{ll}
\lambda_q n_F(|p_0|), & p_0>0, \\
1-\lambda_qn_F(|p_0|)], & p_0<0
\end{array}
\right .
\end{equation}

it was shown, that screening mass is
$$
\bar m^2=\frac{g^2 T^2}{9}\left (\lambda_g+\frac{\lambda_q}{2}\right )
$$

and emission rate is given by formula
\begin{equation}
\label{ratenoneq}
E\frac{dN}{d^4xd^3 p}=\sum\limits_ {u,d,...} e_q^2
\frac {\alpha \alpha_s}{2\pi^2}\lambda_q T^2
e^{-E/T} \left [\frac 23 \left
(\lambda_g+\frac{\lambda_q}{2}\right )\log \left
(\frac{2ET}{\bar m^2_q}\right ) +\frac 4{\pi^2}
C(E,T,\lambda_q,\lambda_g)\right ],
\end{equation}
where

\begin{eqnarray}
C(E,T,\lambda_q,\lambda_g)& = &
\lambda_q\left [-1+\left(1-\frac{\pi^2}6\right )\gamma + \left
(1-\frac{\pi^2}{12}\right )\ln\frac ET+\zeta_-\right ]
+\lambda_g\left [\frac 12+\left(1-\frac{\pi^2}3\right )\gamma + \left
(1-\frac{\pi^2}{6}\right )\ln\frac ET-\zeta_+\right ] \nonumber \\
&&+\lambda_q\lambda_g\left [\frac 12-\frac{\pi^2}{8}+
\left(\frac{\pi^2}4-2\right )\left (\gamma + \ln\frac ET\right ) +\frac
32\zeta^\prime(2)+\frac{\pi^2}{12}\ln 2+\zeta_+-\zeta_-\right ] \nonumber
\end{eqnarray}
and
$\gamma=0.577...$ -- Euler constant,
$\zeta_+=\sum\limits_{n=2}^{\infty}\frac{1}{n^2}\ln(n-1)\approx 0.67$,
$\zeta_-=\sum\limits_{n=2}^{\infty}\frac{(-1)^n}{n^2}\ln(n-1)\approx -0.04$
and $\zeta^\prime$ -- derivative of Rieman $\zeta$-function
($\zeta^\prime(2)=-2.404$).

In the case $\lambda_q=1$ and $\lambda_g=1$ this formula coincides with
(\ref{kapli}). Emission rates, evaluated with this formula are shown on fig.\
\ref{rate-QGP-fig} by dotted line ($\lambda_q=0.5,\, \lambda_g=1$) and solid
line ($\lambda_q=0.5,\, \lambda_g=0.5$). As one can see, emission rate is
very sensitive to the undersaturation of charged constituents, while
undersaturations of gluons slightly vary emission rates.

There are three main sources of uncertainties of emission rate. First --
contributions from higher order in $\alpha_s$ processes. As we have seen,
taking into account bremsstrahlung and annihilation with scattering on third
particle in addition to Compton scattering and annihilation result in
increasing of emission rate up to order of magnitude. So one can not be sure
that inclusion of processes of the next order in $\alpha_s$ does not result
in similar changes of emission rate. Second -- all considered evaluations are
performed in the framework of Braaten - Pisarski Hard Thermal Loop technique,
applicable for the case $g\ll 1$. To evaluate yield of thermal photons in the
heavy ion collisions we have to extend these results for the case
$\alpha_s\sim 0.4$ e.g. $g\sim 2$. Third -- in our evaluations we consider
region $E_\gamma \gg T$ -- the right tail of the thermal distribution. It is
well known, that thermal distribution on the tails forms much more slower,
than in the region $E\sim T$, where 1--2 collisions are sufficient for
establishing of thermal equilibrium.  So, if temperature is rapidly changes
with time, then long tail is not able to form, and it is the region, where
this effect is important. In contrast to the first two sources, the last
effect results in changing of the shape of the distribution and can even
result in nonexponential distribution.

%========================================================================

\subsection{Emission rate from hadronic gas}

\begin{figure}[ht]
\vspace*{9.cm}
\includegraphics{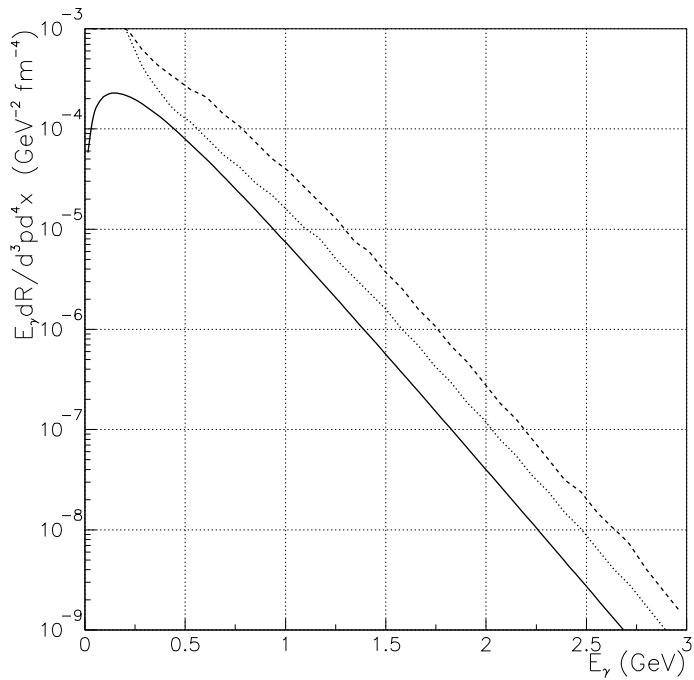}
\caption{Emission rates of thermal photons from hadronic gas at $T=180\,
         MeV$, evaluated with emission rate (\protect\ref{kapli}) (solid
         line), with accounting of $a_1$ resonance
         (\protect\ref{rate-a1-par}) (dotted line), and with accounting of
         in-medium effects (dashed line).}
\label{rate-hadr-fig}
\end{figure}

Thermal photons can be emitted from hadronic gas as a result of reactions
$\pi\pi\to \rho\gamma$, $\pi\rho\to \pi\gamma$, $\rho\to \pi\pi\gamma$,
$\pi\pi\to\eta\gamma$ etc. In the paper \cite{rate-htl} it was shown, that
processes $\pi\rho\to \pi\gamma$ plays the dominant role in emission of
photons with energy larger than $\sim 0.7 \, GeV$. In addition, it was shown,
that photon emission of the hadronic matter is very similar to the photon
emission of QGP (including only Compton scattering and annihilation) at the
same temperature. So one can parametrize photon emission rate from hadronic
gas in the form (\ref{kapli}). Emission rate of hadronic gas, evaluated at
$T=180\, MeV$ in accordance with this parametrization is shown on fig.\
\ref{rate-hadr-fig} by solid line.

Later it was shown \cite{rate-A1}, that contribution of $A_1$ resonance into
process $\pi\rho\to A_1\to\pi\gamma$ is important: it increase emission
rate approximately twice. There was proposed following parametrization of
emission rate with accounting of contribution of $A_1$ resonance:

\begin{equation}
  \label{rate-a1-par}
 E\frac{dR}{d^4xd^3 p}=2.4\, T^{2.15} \exp\left( -1/(1.35\,T\,E)^{0.77}
 - E/T\right )  \quad (fm ^{-4} GeV ^{-2})
\end{equation}

Considered emission rates are evaluated with vacuum masses and widths of
hadrons.  However, these properties change in medium, so that evaluated
emission rates should change too. In the paper \cite{rate-hadr-med} an
attempt was made to evaluate photon emission rate with accounting of such
effects. Effective Lagrangian approach was used to evaluate in-medium masses
and widths. It was shown, that due to decreasing of vector meson masses yield
of thermal photons increases. These predictions for hadronic gas at $T=180\,
MeV$ are shown on the fig.\ \ref{rate-hadr-fig} by dashed line.

From one hand, situations with emission rate from hadronic gas is almost so
shaky as with emission rate from QGP -- as far as there is no theory of
strong couplings. From the other hand, one can normalize coupling
constants and other hadronic properties to experiment as it was done in
\cite{rate-A1}, what can not be done in the case of QGP. In addition to this,
evaluated emission rates suppose chemical equilibrium. However, it may not be
established due to various reasons: fashion of hadronization of QGP, fast
cooling of hadronic gas, fly out of hadrons from QGP surface \cite{we} and
hadronic gas, etc. As it was shown e.g. in \cite{rate-htl}, main contribution
to emission rate comes from pion-pion and pion-rho scatterings.  So, the most
important for us is oversaturation of pions and rho due to fast cooling. One
can estimate this oversaturation as ratio $N(T_c)/N(T_f) \sim (T_c/T_f)^3$,
where $T_c$ and $T_f$ -- transition and freeze-out temperatures
correspondingly. So one could expect, that emission rate increase in the same
ratio. So we estimate uncertainty in emission rate from hadronic gas within
factor $\sim 3$ due to in-medium effects and factor $\sim 3$ due to
oversaturation of pions and rho in hadronic gas.

%=======================================================================

\subsection{Yields of thermal photons}

To evaluate yield of thermal photons in the nucleus-nucleus collision one
should convolute emission rate with space-time evolution of the system:
\begin{equation}
E_\gamma\frac{dN}{d^3p}=\int E_\gamma \frac{dR}{d^3p}(T)\, dV dt
\end{equation}

Below we consider several models of evolution of hot matter and evaluate
yield of thermal photons in these models.

First of all, on the example of very simple 1+1 (1 space+ time) Bjorken
hydrodynamic model \cite{Bjorken} we show main phenomena: dependence on
initial conditions, transition temperature etc., and then compare predictions
of Bjorken model with 1+1 Landau hydrodynamics and more advanced model such
as (2+1) Bjorken model.

In the 1+1 Bjorken hydrodynamic model of evolution of heavy ion collision it
is assumed, that hot matter, created in the beginning of collision, expands
only longitudinally (along collision axis) and longitudinal velocity depends
on longitudinal coordinate in the following way:
$$
u^\mu=\left (\frac{t}{\tau},0,0,\frac{z}{\tau}\right ), \quad
\tau=\sqrt{t^2-z^2}.
$$

This parametrization of velocity result in flat rapidity distribution of
fluid elements, and as a result -- of final hadrons. So, this model is well
applicable only in the midrapidity region. Below we restrict ourselves by
midrapidity region $z=0$, $\tau=t$, because this region is most interesting
from point of view of observation of hot hadronic matter.

As a result of such expansion, the volume of the hot matter increase with time
in the following way:
$$
V=2\,\pi R^2\, t,
$$
where $R$ -- radius of colliding nuclei. If one assumes now, that entropy is
conserved during all the expansion, one can relate temperature and time:
$$
S=s\cdot V=const\cdot \left( \frac{dN}{dy}\right ), \quad
s=4\frac{\pi^2}{90}\,g\, T^3
$$
or
$$
T^3t=\frac{\pi}{\zeta(3) R^2 g} \frac{dN}{dy},
$$
where $T$ -- temperature, $t$ -- time, $g$ -- degeneracy, $dN/dy$ --
multiplicity of final (massless) hadrons (bosons) and
$\zeta(3)=1.202$. In the simplest version of Bjorken hydrodynamics
we have $g_{QGP}=16+\frac 78 24$ for degeneracy of QGP and $g_{hadr}=3$ for
hadronic gas with dominating pions.

Now, choosing from some considerations e.g. from cascade model multiplicity
$dN/dy$ and initial time (initial temperature) -- the time (temperature), at
which we already can assume thermal equilibrium, we find initial conditions.
Then, fixing transition temperature $T_c$ and freeze-out temperature $T_f$,
we completely define this model.

Having emission rate of photons (e.g. (\ref{kapli})) and description of
evolution, one can evaluate the total number of thermal photons as function
of model parameters:

\begin{equation}
\label{tot-num}
N_\gamma \sim \left(\frac{dN}{dy}\right )^2
\left \{\frac{1}{g_h^2T_f^2}-
\frac 23 \frac{g_{QGP}^2-g_h^2}{g_{QGP}^2g_h^2}\frac 1{T_c^2}-
\frac 1{g_{QGP}^2}\frac 1{T_{in}^2}    \right \} .
\end{equation}

As one can easily see, the total number of thermal photons slightly depends
on the initial temperature, but strongly -- on freeze-out one. This is due to
emission of very soft photons during prolongated evolution in the case of
lower $T_f$. However, we are interested in the hard part of the spectra, so
that this quantity is not appropriate for our purposes. More informative
would be $p_t$ distribution of thermal photons.

Let us now fix basic set of parameters of the model:
\begin{equation}
\label{param}
dN/dy=12000, \quad T_{in}=1\, GeV,\quad T_c=160\, MeV,\quad T_f=100\, MeV,
\end{equation}
and look, how $p_t$ distribution of direct photons depends on variation of
model parameters.

\begin{figure}[h]
\vspace*{9.5cm}
\includegraphics{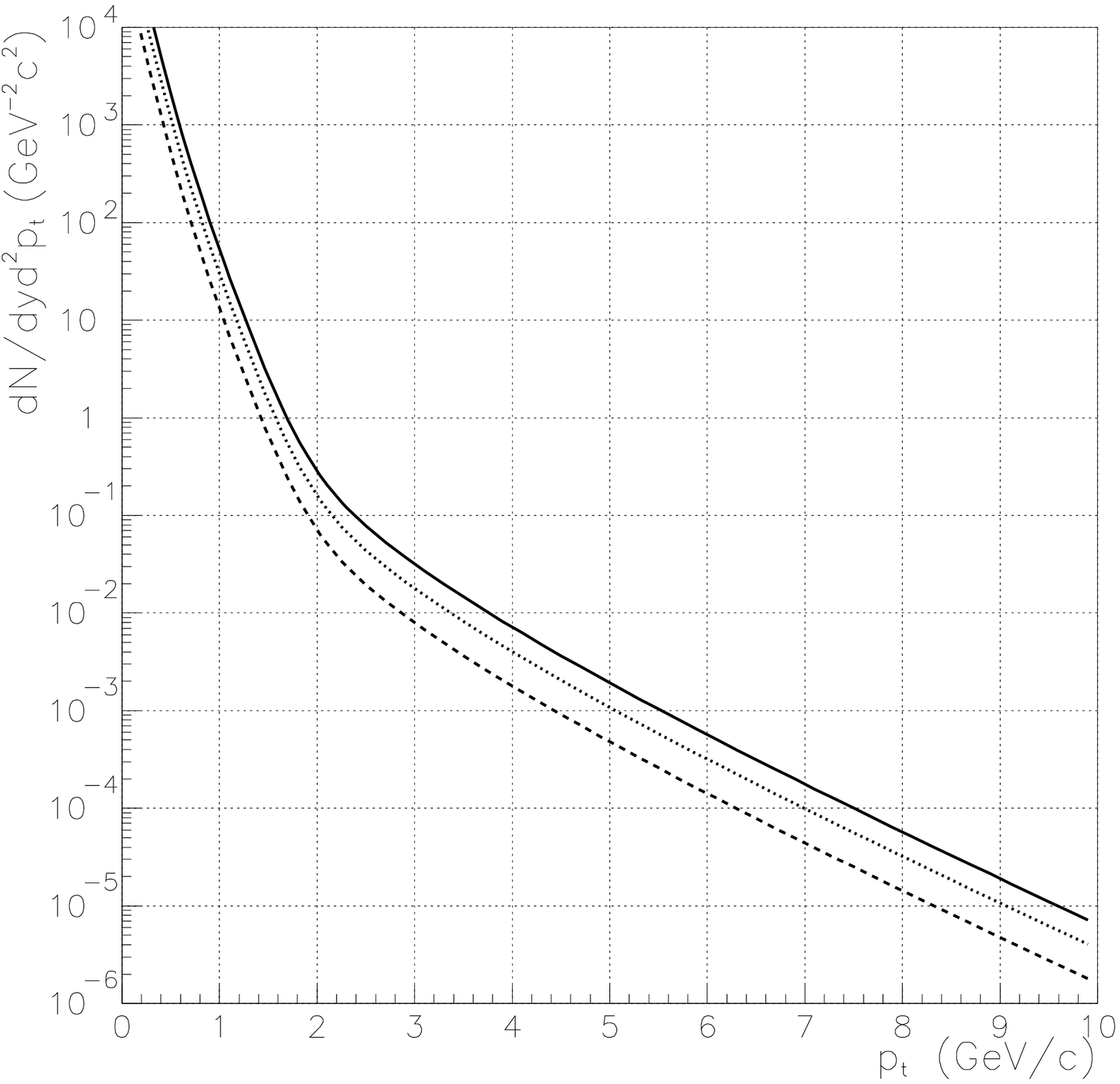}
\includegraphics{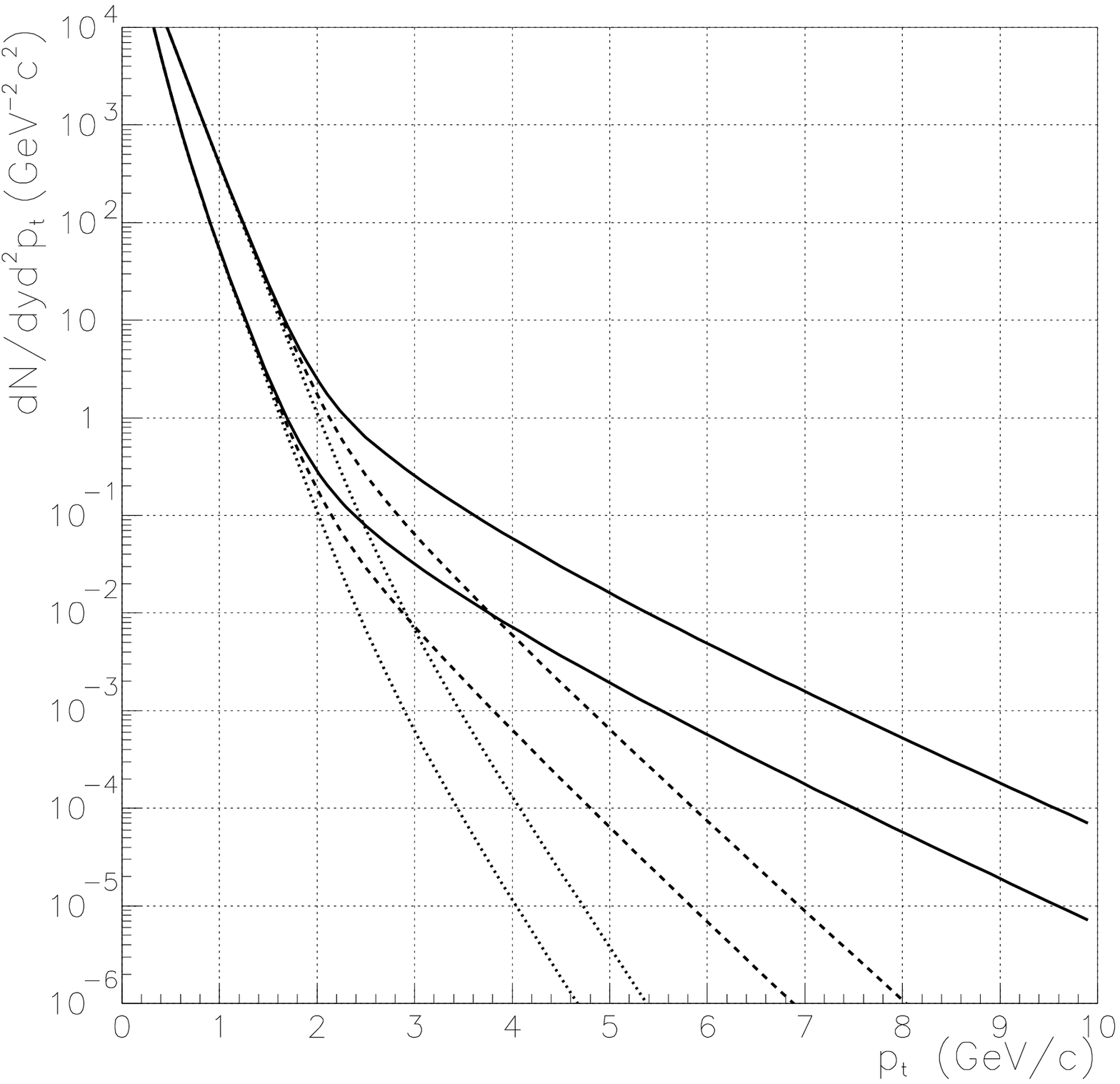}
\caption{Yield of thermal photons in $Pb+Pb$ collision at LHC
         energy within 1+1 Bjorken hydrodynamics. Left plot: yield evaluated
         using emission rate (\protect\ref{kapli}) both for QGP and hadronic
         phase and parameters (\protect\ref{param}) for $dN/dy=12000$ (solid
         line), $dN/dy=9000$ (dotted line) and $dN/dy=6000$ (dashed line).
         Right plot:  Yield, evaluated with parameters (\protect\ref{param})
         and initial temperatures:  $T_{in}=1\, GeV$ (solid lines),
         $T_{in}=500\, MeV$ (dashed lines) and $T_{in}=300\, MeV$ (dotted
         lines). Upper three curves evaluated using emission rates
         (\protect\ref{twoloop-brems}),(\protect\ref{twoloop-ann}) for QGP
         phase, and (\protect\ref{rate-a1-par}) for hadronic phase, bottom
         three curves -- using emission rate (\protect\ref{kapli}) both for
         QGP and hadronic phase.
         }
\label{1dbjor-np}
\end{figure}

To begin with, let us vary multiplicity -- see fig.\ \ref{1dbjor-np}. With
fixed initial temperature variation of $dN/dy$ means variation of initial
time, and thus all times of the evolution: all times change in the same
scale. As one can expect from (\ref{tot-num}), the number of photons varies
as squared multiplicity, while the shape of distribution does not changes.
However, this effect holds only for 1+1 Bjorken hydrodynamics. In more
realistic models the dependence of $dN^\gamma/dy$ on the total multiplicity
is weaker. For example in the 2+1 bjorken hydrodynamics $N^\gamma\sim
(dN/dy)^{1.2}$ \cite{Ngam-dNdy}. This can be understood in the following way.
The number of thermal photons is proportional to the number of charged
particles times the number of collisions, suffered by each particle:

$$
N_\gamma\sim N_{ch} \nu.
$$

For the case of long lived system each particle can collide to all other
particles, so $\nu\sim N_{ch}$. However, if system expands, or has final
dimensions comparable or even less then free path length, then the number of
collisions
$$
\nu\sim \frac D\lambda
$$
where $D$ - dimensions of the system. In the case of 1+1 Bjorken hydrodynamics
increasing of $dN/dy$ result in increasing of longitudinal dimensions of the
system at fixed transverse sizes, so $D\sim dN/dy$
$$
N_\gamma\sim (dN/dy)^2.
$$

If we allow transverse expansion, then $D\sim (dN/dy)^{1/3}$ and
$$
N_\gamma\sim (dN/dy)^{4/3},
$$
result very close to obtained in 2+1 Bjorken hydrodynamics.

As a next step we vary initial temperature -- see fig.\ \ref{1dbjor-np},
right plot.  With fixed $dN^\pi/dy$ this means that we cut off
initial stages of evolution: in the case $T_{in}=500\, MeV$ we skip
time $\tau=0.15\to 1\, fm/c$ with respect to the case $T_{in}=1\, GeV$ and
in the case $T_{in}=300\, MeV$ -- $\tau=0.15 \to 4.6\, fm/c$. So, from fig.\
\ref{1dbjor-np} one can find what part of spectrum is formed at each stage of
collision. We obtain very natural result: the harder part of spectrum is
populated due to emission from the hottest stage of collision. In addition, on
this figure we compare yields of thermal photons, evaluated using simplest
emission rate (\ref{kapli}) (three lowest curves) and emission rates,
accounting second order QCD corrections (\ref{twoloop-brems},\ref{twoloop-ann})
and $a_1$ contribution (\ref{rate-a1-par}) (three upper curves). In the
latter case yield of thermal photons is approximately order of magnitude
larger.

\begin{figure}[ht]
\vspace*{9.5cm}
\includegraphics{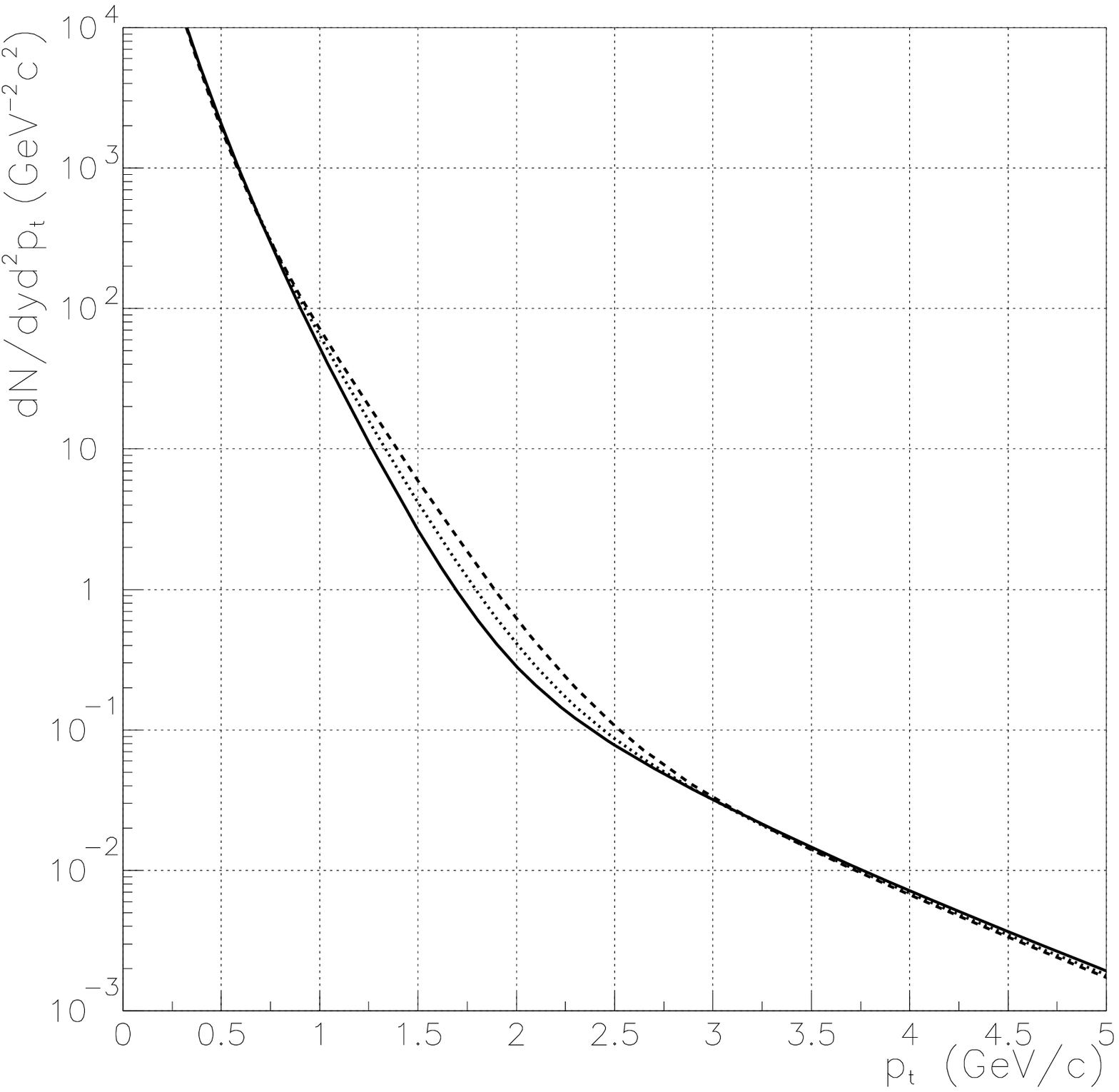}
\includegraphics{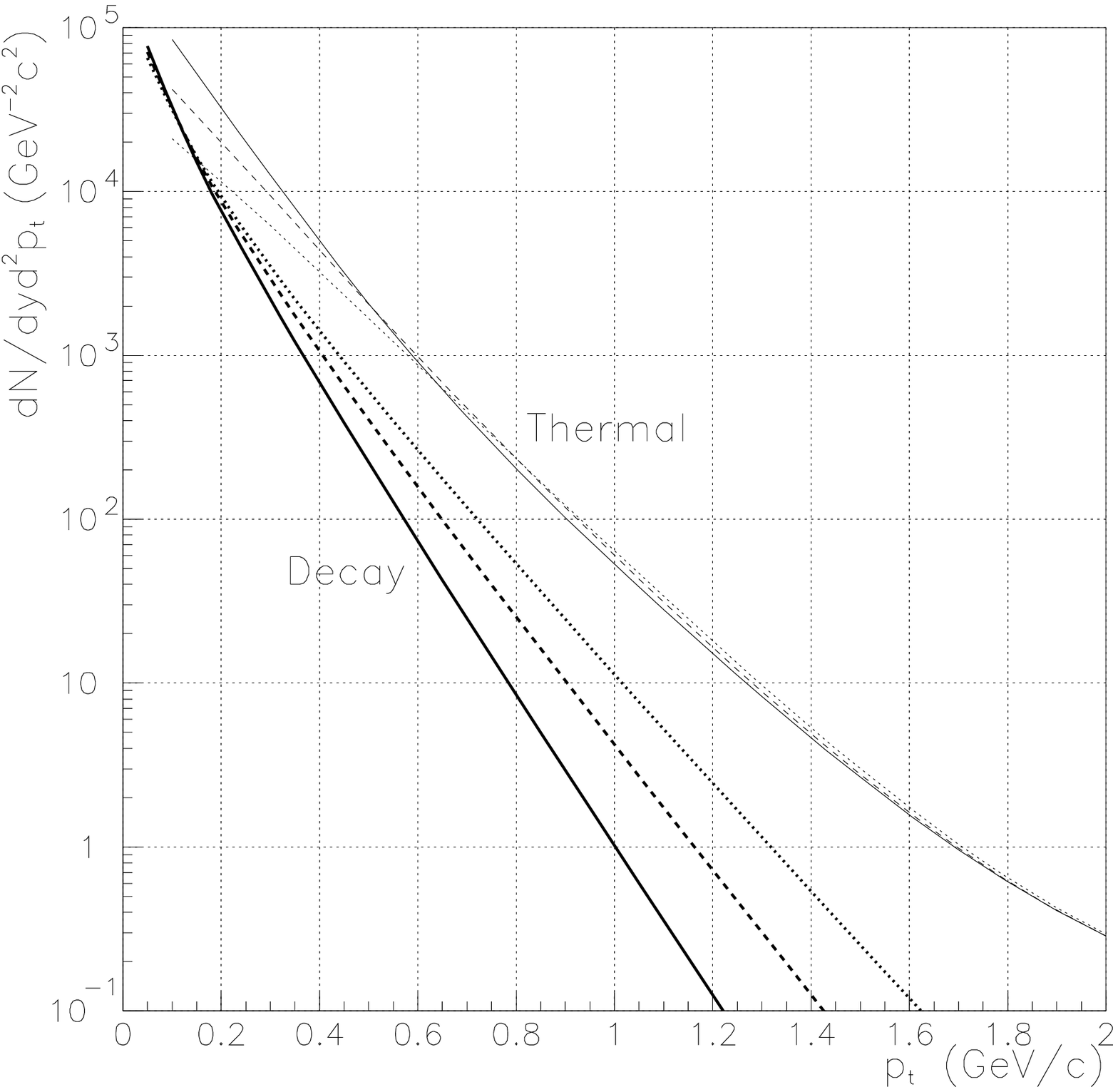}
\caption{Left plot: The same as fig.\ \protect\ref{1dbjor-np}, but with
         different transition temperatures $T_c=200\, MeV$ (dashed line),
         $T_c=180\, MeV$ (dotted line) and  $T_c=160\, MeV$ (solid line).
         Right plot: thermal and decay photons evaluated within 1+1
         Bjorken hydrodynamics with different
         freeze-out temperatures: $T_{f}=100\, MeV$ (solid line),
         $T_{freeze}=120\, MeV$ (dashed line) and  $T_c=140\, MeV$
         (dotted line). In all cases thermal photons are evaluated with
         emission rate (\protect\ref{kapli}).
         }
\label{1dbjor-tc}
\end{figure}

Next step is variation of the transition temperature $T_c$ -- see fig.\
\ref{1dbjor-tc}, left plot. As we see below, mixed phase lives much more
longer, than QGP phase, and thus a lot of thermal photons are emitted with
$T=T_c$, increasing yield of medium energy direct photons with increasing of
$T_c$.  Lifetime of mixed phase is determined by ratio of degeneracies of QGP
phase and hadronic phase. The simplest Equation of State (EOS) of hadronic
gas corresponds to ideal gas of massless pions, so that this ratio is
$37/3\approx 12$ and mixed phase lives a very long time. If one considers
hadronic gas, consisting from all known hadrons with masses below $2\, GeV$,
then this ratio becomes $37/12\approx 3$ and time of life of mixed phase is
significantly reduced, reducing yield of medium energy photons.

And finally we vary freeze-out temperature $T_f$. Its variation affects both
$p_t$ distributions of thermal photons and $p_t$ distributions of final
hadrons, and thus decay photons -- so that we show both decay and
direct photons on the fig.\ \ref{1dbjor-tc}, right plot. From one hand,
decreasing of $T_f$ result in increasing of time of life of hadronic
gas and thus yield of soft thermal photons, from the other hand it
result in decreasing of the temperature of final hadrons, so that ratio
$thermal/decay$ increase.

Comparing figs.\ \ref{1dbjor-np}-\ref{1dbjor-tc}, one can conclude, that
within 1+1 Bjorken hydrodynamic model hard part of spectrum of thermal
photons $p_t> 2\, GeV/c$ is governed by emission from the hottest region
(QGP), variation of transition temperature vary $p_t$ distribution at
$1<p_t<2.5\, GeV/c$, while the softest part of the spectrum $p_t<0.5\, GeV/c$
depends on the freeze-out temperature.

It is remarkable (see fig.\ \ref{1dbjor-tc}, right plot), that within 1+1
Bjorken hydrodynamics direct photons dominate over decay ones
practically at all energies. We obtain this unphysical result because we
neglect transverse expansion. In the case of thermal photons absence of
transverse expansion result in extremely large times of evolution: e.g. for
our set of parameters (\ref{param}) we find

\begin{equation}
\label{times-1d}
\tau_0=0.15\,fm/c,\quad \tau_{qgp}=12\,fm/c,\quad\tau_{mixed}=137\,fm/c,\quad
\tau_{f}=560\, fm/c,
\end{equation}
where $\tau_0$ -- time of thermalization, $\tau_{qgp}$ -- time of
disappearing of QGP, $\tau_{mixed}$ -- time of disappearing of mixed phase,
$\tau_f$ -- time of freeze-out. So that, emission of photons from the mixed
phase and especially from the hadronic phase is strongly overestimated in
comparison with the models with transverse expansion.

In the case of decay photons, absence of transverse expansion result in lower
effective temperature, and smaller contribution with respect to thermal
photons.

However, this does not mean, that 1+1 Bjorken model is useless: as we see
below, it makes reasonable predictions for hard part $p_t> 2\, GeV/c$ of the
spectrum of thermal photons, emitted on the stage, where transverse expansion
is not very important.

Bjorken model of evolution is essentially nonhydrodynamic, i.e.  it is not
solution of the hydrodynamic equations. So, it is interesting to compare its
predictions with really hydrodynamic model e.g. 1+1 Landau hydrodynamics
\cite{landau}. Within this model it is assumed, that just after penetration
of colliding nuclei through each other a thermalized matter is created,
having form of Lorentz compressed disk. The part of the total energy, deposed
in this disk is model parameter -- coefficient of inelasticity $K$. Being
formed, disk expands along beam axis in accordance with one dimensional
hydrodynamic equations.  Initially disk has size along beam axis:

\begin{equation}
\label{lan-width}
\Delta=\frac{4 m_p R}{3 K\sqrt{s}}\frac{c^2_s}{1+c^2_s},
\end{equation}
where $m_p$ is the proton mass, $K$ -- coefficient of inelasticity $K\sim
0.5$, $R$ -- radius of nucleus and $c_s^2$ -- speed of sound in the created
matter. For $Pb+Pb$ collision ($R=6.5\, fm$) at LHC energy ($\sqrt{s}=6300\,
A\cdot GeV$) one finds ($K=0.5, c_s^2=1/3$):

$$
\Delta\approx 0.0025\, fm.
$$

Initial temperature can be evaluated from the condition
$$
2\pi R^2 \Delta \cdot\varepsilon = K \sqrt{s}\cdot A,
$$
where $\varepsilon$-energy density. Assuming QGP consisting of two quark
flavors and gluons, one finds for $Pb+Pb$ collisions at LHC energy:
$$
T_{in}=5.1\, GeV.
$$

\begin{figure}[ht]
\vspace*{9.5cm}
\includegraphics{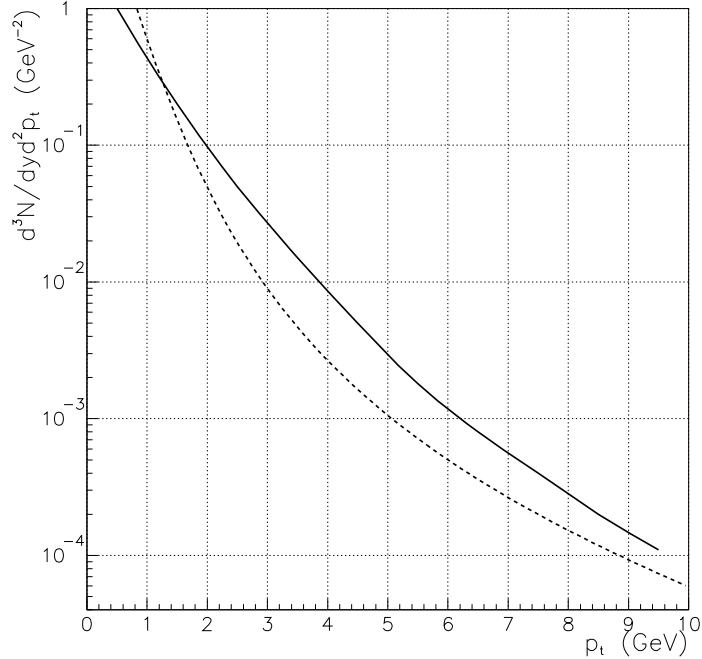}
\caption{Yield of thermal photons, emitted from QGP phase in
         $Pb+Pb$ collision at LHC energy, evaluated within Landau
         hydrodynamic model (solid line), and 1+1 Bjorken hydrodynamics
         with $T_0=5.1\, GeV$, $T_{c}=220\, MeV$ and
         $\tau_{in}=0.0025\, fm/c$ (dashed line).}
\label{land-bjor}
\end{figure}

Result of evaluation of the yield of thermal photons from QGP phase only
($T_c=220\, MeV$) within Landau model taken from \cite{landau} is presented
on the fig.\ \ref{land-bjor} by solid line. Dashed line corresponds to the
evaluation within 1+1 Bjorken model with the same initial conditions:
$T_{in}=5.1\, GeV$, $T_c=220\, MeV$ and $\tau_{in}=0.0025\, fm/c$. This
conditions correspond to multiplicity $dN^\pi/dy=5300$. As one can see, lines
goes rather close to each other, though there is some difference due to
different rapidity distributions and, as a consequence, different dependence
$T(\tau)$. Scaling expansion result in faster cooling in the beginning and
slower -- on the later stages. Nevertheless, this result demonstrate, that
use of scaling longitudinal expansion instead of hydrodynamic one is not
critical for our purposes.

Taking into account transverse expansion alters significantly all evolution of
the system, and thus spectra of the thermal photons. To include transverse
expansion we consider 2+1 Bjorken hydrodynamics. This means, that we assume
Bjorken scaling expansion along beam axis, while for expansion in transverse
direction we solve hydrodynamic equations. So, we parametrize 4-velocity in
the form:
$$
u^\mu=\gamma (\frac{t}{\tau},\vec v_r,\frac{z}{\tau})\quad
\tau =\sqrt{t^2-z^2}, \quad
\gamma =\left( 1-v^2\right) ^{-\frac 12},
$$
and solve hydrodynamic equations

\begin{equation}
\label{dT}\partial _\mu
T^{\mu \nu }=0,\qquad
T^{\mu \nu }(x)=
\left( \varepsilon (x)+p(x)\right) \cdot u^\mu u^\nu -g^{\mu \nu }\cdot p(x),
\end{equation}
where $T^{\mu \nu }$ -- energy-momentum tensor of ideal fluid, $\varepsilon
(x)$ and $p(x)$ -- local energy density and pressure correspondingly, $u^\mu
$ -- 4-velocity and $g^{\mu \nu }$ metric tensor ($g=diag(1,-1,-1,-1)$). In
the case of cylindrically symmetrical transverse expansion and longitudinal
scaling expansion one can rewrite (\ref{dT}) in the form

\begin{eqnarray}
\label{part}
{\partial _\tau \left( \gamma ^2w\right) +
\partial _r\left( \gamma ^2vw\right) +
\gamma ^2w\tau^{-1} + \gamma ^2vwr^{-1}-\partial _\tau p=0}  \qquad\,\,
\nonumber \\
{\partial _\tau \left( \gamma
^2vw\right) +\partial _r\left( \gamma ^2v^2w\right) +
\gamma ^2vw\tau^{-1} +\gamma ^2v^2wr^{-1}+\partial
_rp=0},
\end{eqnarray}
where $w=\varepsilon +p$ -- enthalpy density. We solve this equations
numerically.

As in the one dimensional case we fix basic set of parameters
\begin{equation}
\label{set-2dim}
dN/dy=12000,\quad T_{in}=1\, GeV,\quad T_c=160\, MeV,\quad T_f=140\, MeV.
\end{equation}

\begin{figure}[ht]
\vspace*{9.5cm}
\includegraphics{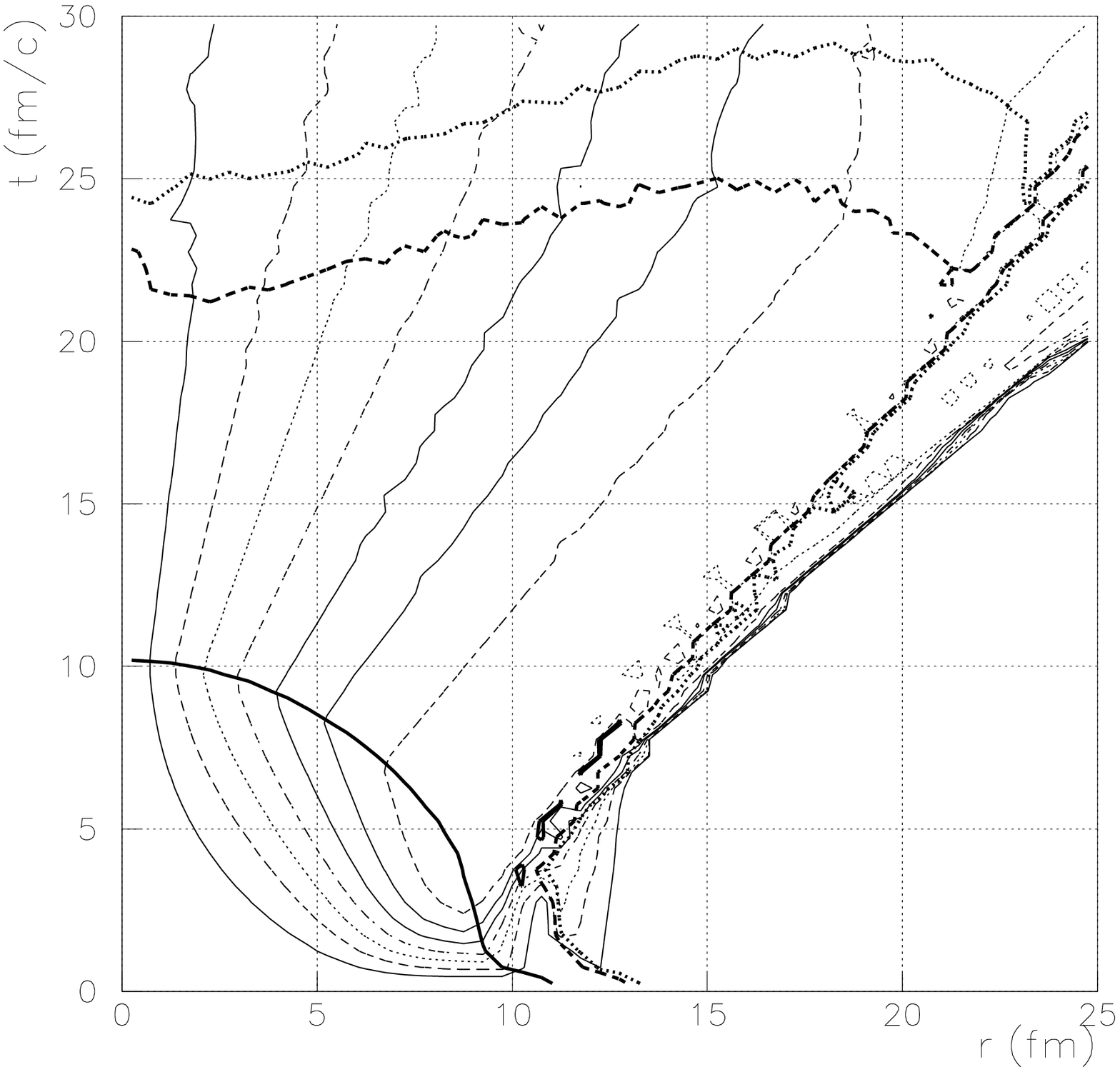}
\includegraphics{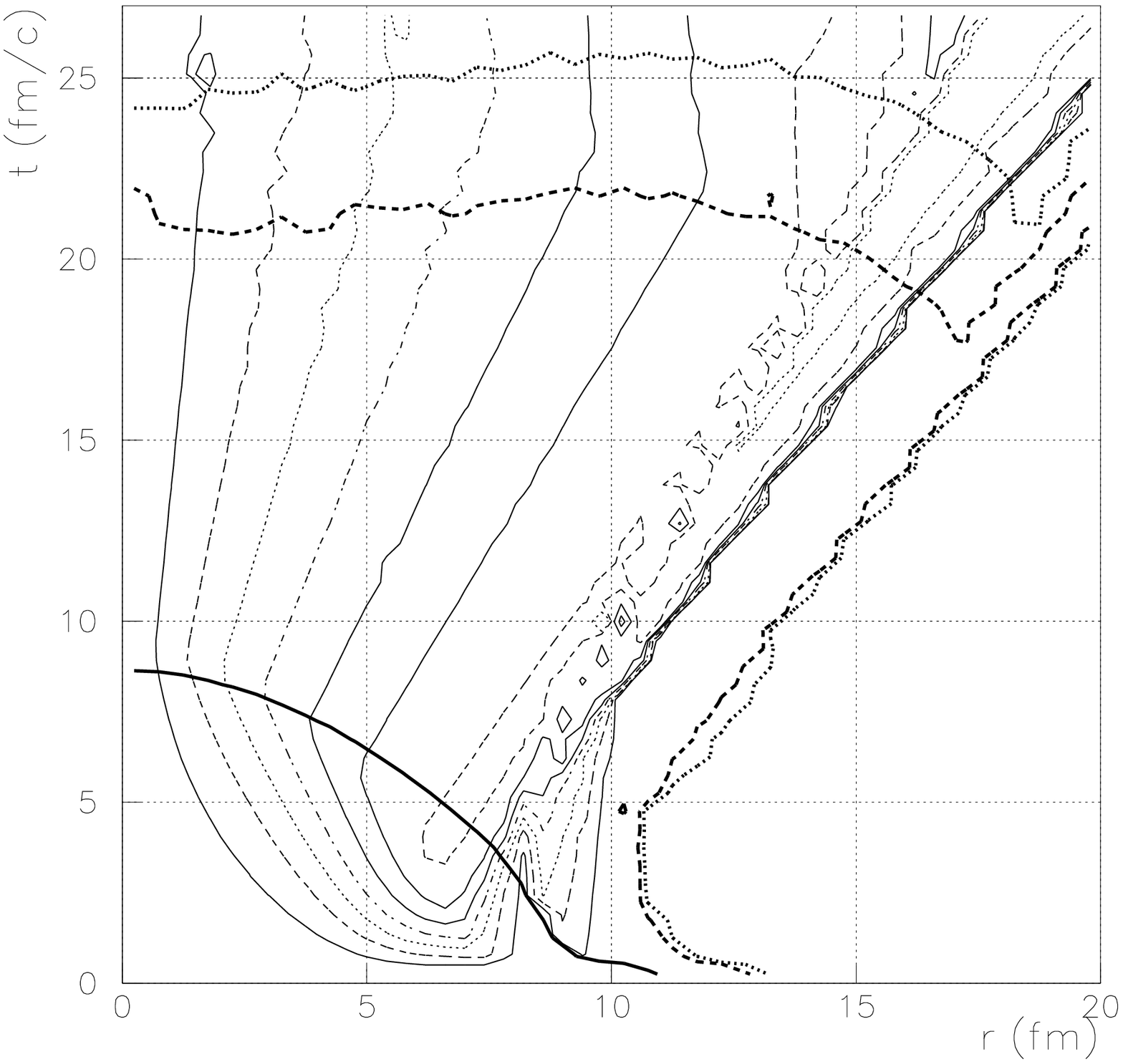}
\caption{Evolution of the hot matter, calculated at LHC energy for
         $dN/dy=12000$ left plot, and $dN/dy=6000$ right plot.  Constant
         velocity levels (thin lines) are drown at 0.1,0.2,... and constant
         energy density levels, are drawn at the boundary of QGP phase (thick
         solid line), at the boundary of mixed phase (thick dashed line) and
         freeze-out surface (thick dotted line).
        }
\label{contour12}
\end{figure}

In contrast to the 1+1 dimensional case, where all time scales are given by
initial time (at fixed initial temperature), in the 2+1 dimensional case we
have two completely different time scales: one of longitudinal expansion -- the
same as in 1+1 dimensional case and one of transverse expansion, determined
by velocity of sound and dimensions of the system. So that, picture of
expansion become much more complicated. Typical evolution of hot matter in
2+1 dimensional case is shown on the fig.\ \ref{contour12} -- this is result
of evaluations with parameters, listed in (\ref{set-2dim}) (left plot) and
$dN/dy=6000,\, T_{in}=1\, GeV,\, T_c=160\, MeV,\, T_f=140\, MeV$ (right
plot). Initially system has zero transverse collective velocity, than due to
acceleration in the QGP phase matter possesses velocity more then $0.7\cdot
c$. In the mixed phase pressure gradient is absent and velocities
are constant (velocity levels are straight), and after the end of the mixed
phase during hadronic gas expansion matter begins to accelerate further
(slopes of velocity levels changed).

In contrast to one dimensional case, times of life of mixed and hadronic
phases is much smaller:

\begin{equation}
\label{times-2d12}
\tau_0=0.11\,fm/c,\quad \tau_{qgp}=10\,fm/c,\quad\tau_{mixed}=25\,fm/c,\quad
\tau_{f}=30\, fm/c.
\end{equation}
in the case of $dN/dy=12000$, and
\begin{equation}
\label{times-2d6}
\tau_0=0.06\,fm/c,\quad \tau_{qgp}=8\,fm/c,\quad\tau_{mixed}=22\,fm/c,\quad
\tau_{f}=25\, fm/c.
\end{equation}
in the case of $dN/dy=6000$.

In the case of larger multiplicity initial time is twice larger, so that in
the beginning stage system cools twice slower, and more energy is spent to
the developing of the radial velocity. Thus we find higher transverse
collective velocity in the case of higher multiplicity. However, later stages
($t > 5\, fm/c$) are governed mainly by radial expansion, so that life times
of QGP, mixed phase and hadronic gas are very similar on both cases.

There are two important consequences of introduction of transverse
expansion. First, -- times of life are changed: time of life of QGP is
changed slightly, but time of life of mixed phase and hadronic gas are
reduced more then order of magnitude compared to 1+1 dimensional case.
Second, -- hot matter poses significant radial velocity, which affect spectrum
of thermal photons and even in the more degree -- spectrum of decay
photons.

\begin{figure}[ht]
\vspace*{9.5cm}
\includegraphics{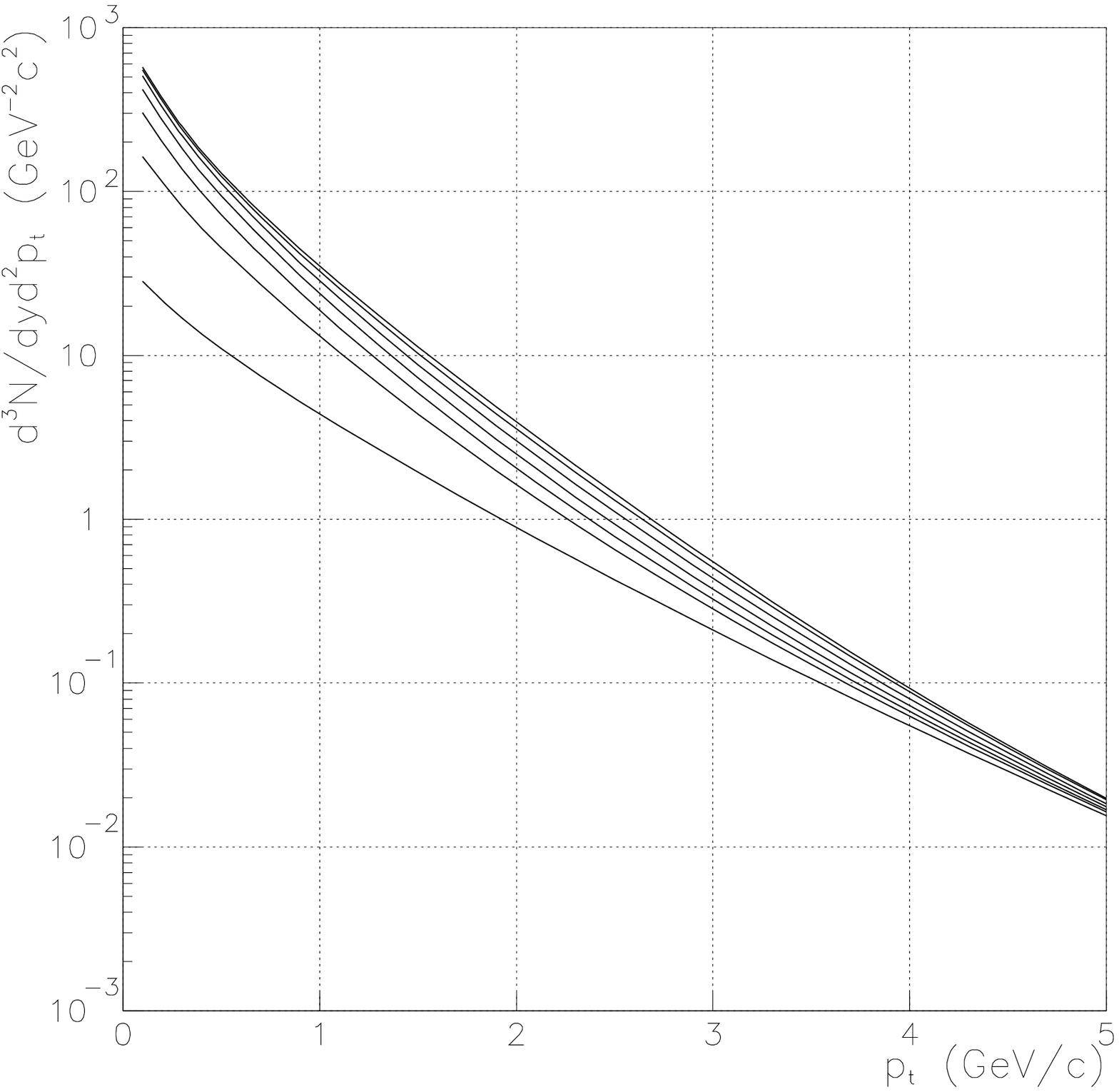}
\includegraphics{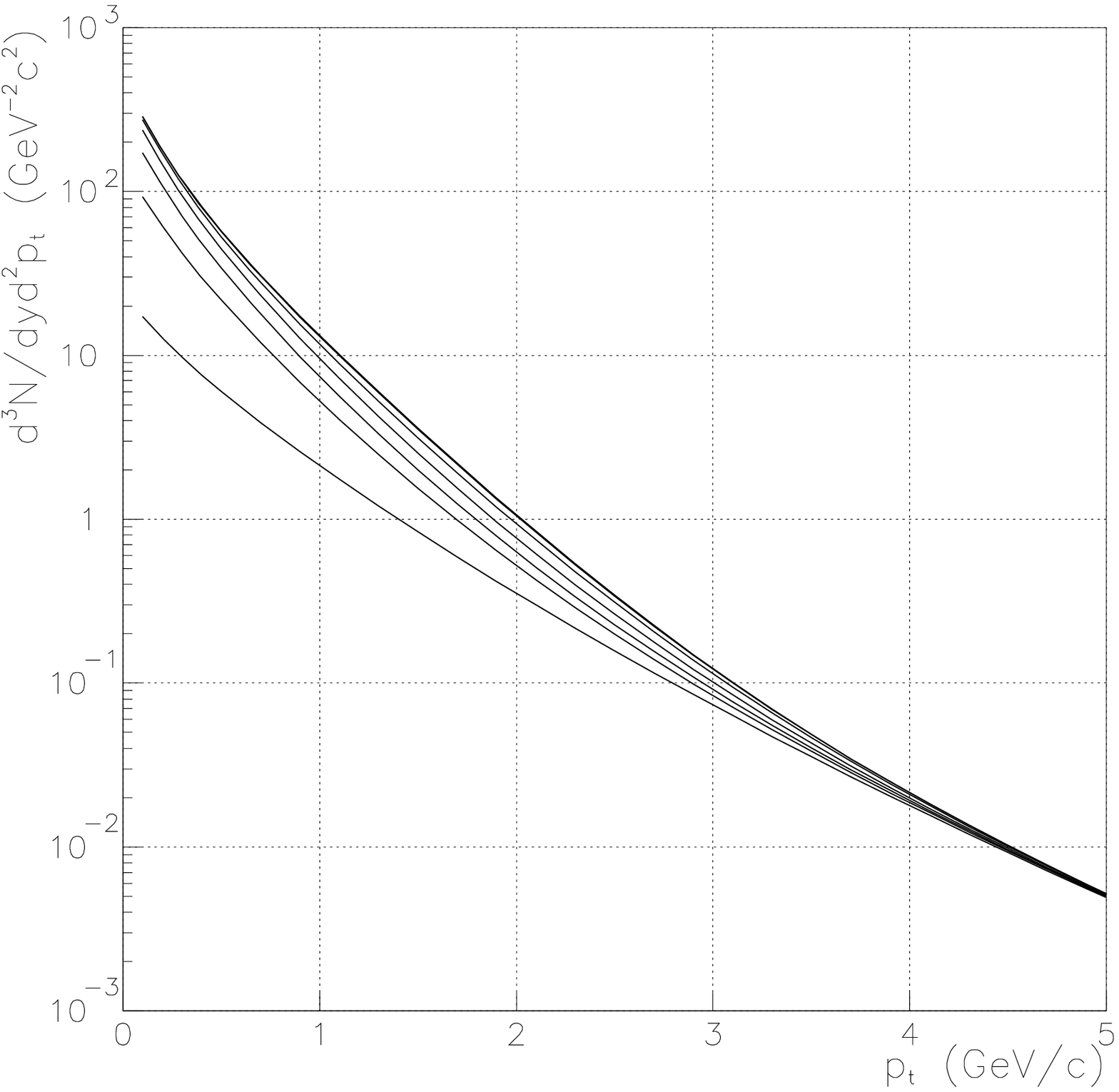}
\caption{Contributions of different times (from down to up 0-1, 0-5, 0-10,
         0-15, 0-20, 0-25 fm/c, total) into the total thermal photon yield,
         evaluated at $dN/dy=12000$ (left) and $dN/dy=6000$ (right).}
\label{times}
\end{figure}

As a result of producing of the large radial velocity the thermal photons,
emitted from hadronic gas contribute into the hard part of the spectrum of
direct photons -- see fig.\ \ref{times}. On these figures contributions of
different times into the total yield of thermal photons are shown. Similar to
one dimensional case, photons emitted at first several $fm/c$ populate hard
part of spectrum. However, due to high collective velocities contributions of
later (cooler) stages into hard region are comparable with emission from the
hottest stage. This effect is better pronounced in the case of higher
multiplicity ($dN/dy=12000$), where collective velocity can reach $v_r\sim
0.9$ and effective temperature

\begin{equation}
\label{teff}
T_{eff} \sim T_f \sqrt{\frac{1+v_r}{1-v_r}}\sim 0.14\, GeV\cdot 4.4 \sim 0.6\,
GeV
\end{equation}
is comparable even with initial temperature.

\begin{figure}[ht]
\vspace*{9.5cm}
\includegraphics{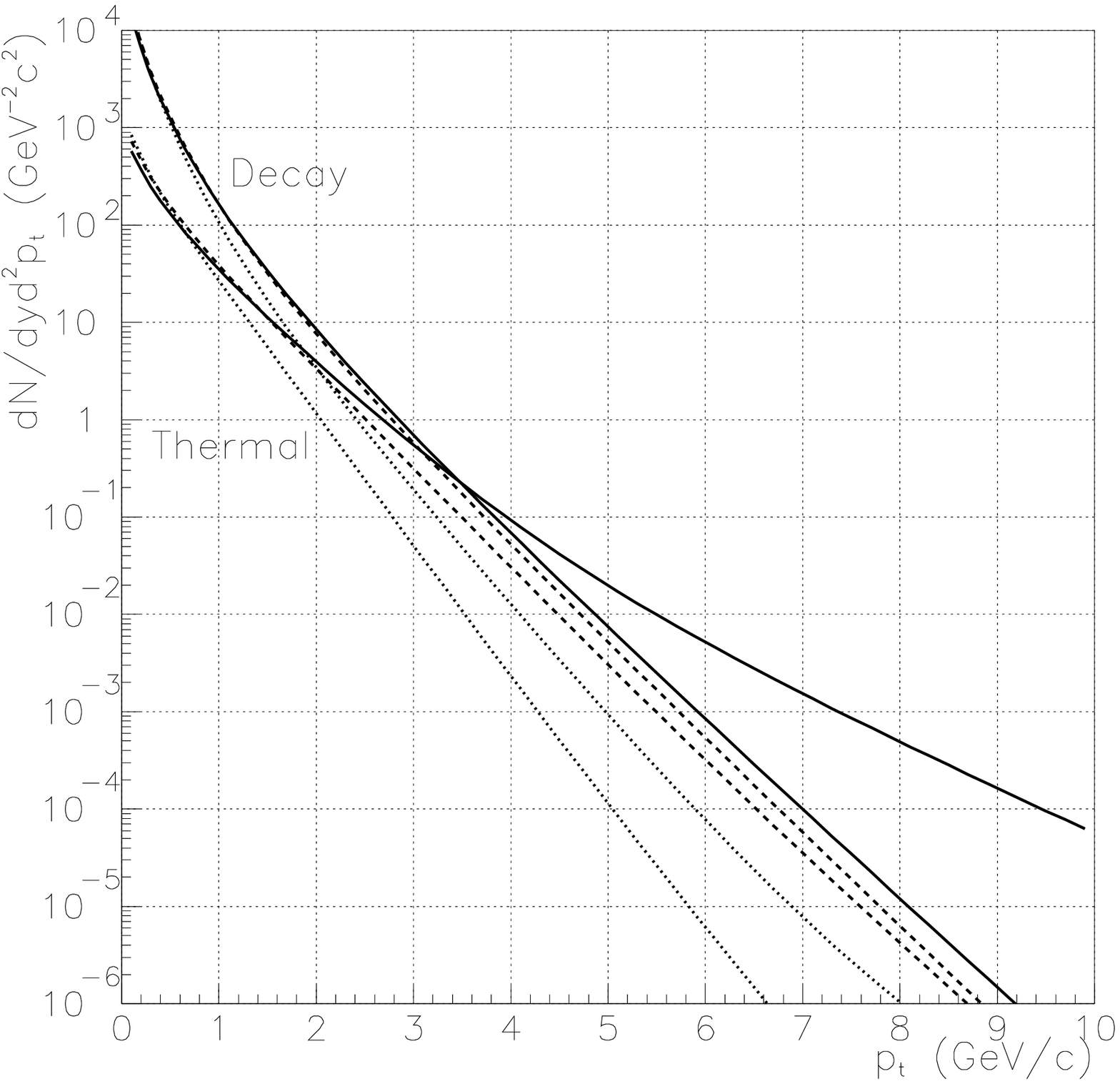}
\includegraphics{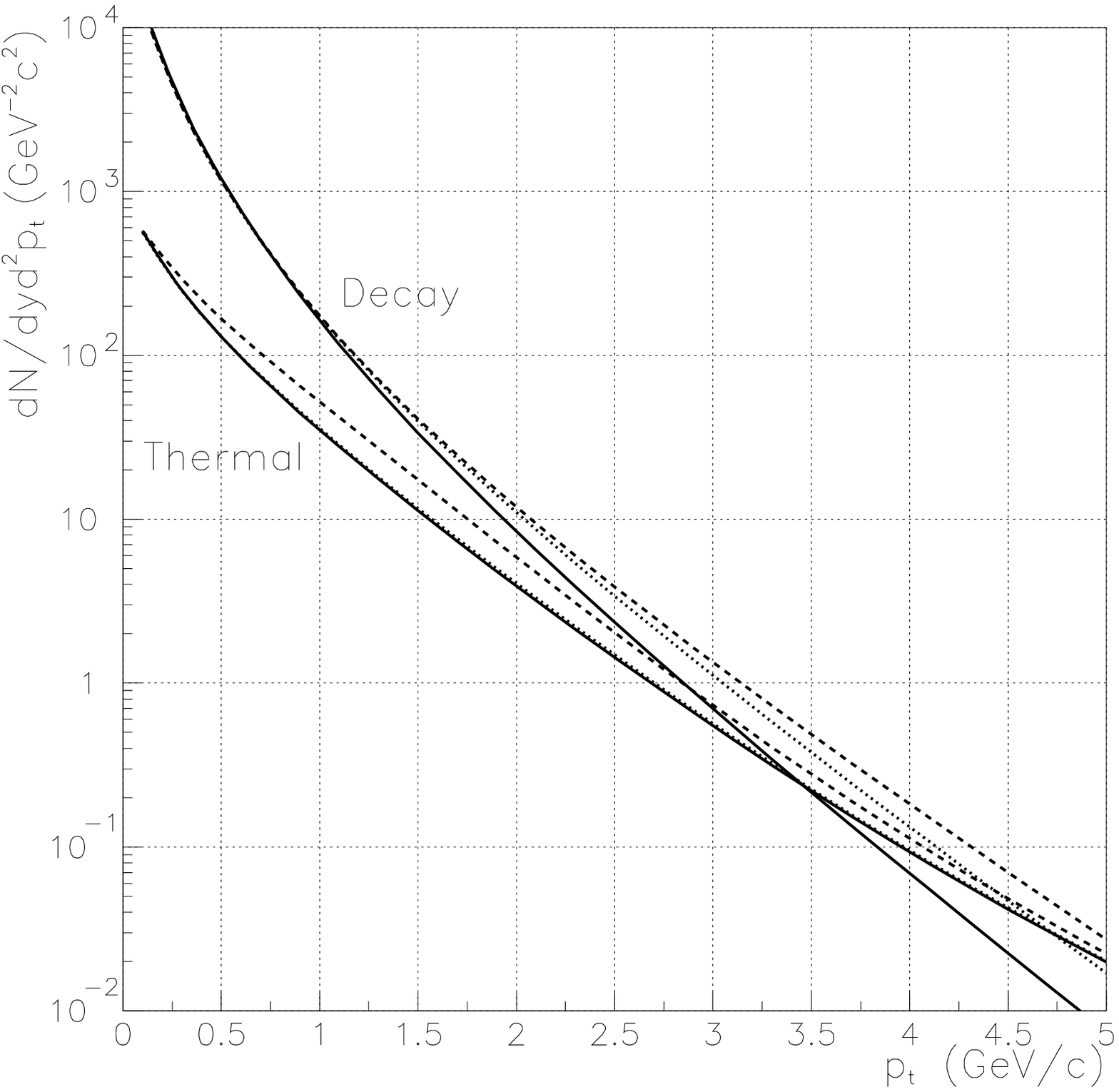}
\caption{Left plot: yield of thermal and decay photons, evaluated within
         2+1 hydrodynamics with set the of parameters (\protect\ref{set-2dim})
         and initial temperature $T_{in}=1\, GeV$ (solid lines),
         $T_{in}=500\, MeV$ (dotted lines) and $T_{in}=300\, MeV$
         (dashed lines).
         Right plot: predictions with the basic set of parameters
         (\protect\ref{set-2dim}) (solid lines), variation of transition
         temperature $T_c=200\, MeV$ (dotted lines) and freeze-out
         temperature $T_f=120\, MeV$ (dashed lines). In all evaluations
         emission rates (\protect\ref{twoloop-brems})
         (\protect\ref{twoloop-ann}) and (\protect\ref{rate-a1-par}) are
         used.}
\label{2dbjor-ti}
\end{figure}

Similar to 1+1 dimensional case we investigate dependence on the parameters
of the model. In particular on initial, transition and freeze-out
temperatures. Yields of thermal and decay photons, evaluated with three
different initial temperatures ($T_{in}=1\, GeV, \, 500 \, MeV$ and $300\,
MeV$), and fixed multiplicity, transition and freeze-out temperatures
(\ref{set-2dim}) are shown on the fig.\ \ref{2dbjor-ti}, left plot. If one
compare spectrum of thermal photons, evaluated within 1+1 (fig.\
\ref{1dbjor-np}, right plot) and 2+1 (fig.\ \ref{2dbjor-ti}) Bjorken
hydrodynamics, one finds, that in the case of $T_{in}=1 \, GeV$ hard parts of
spectra of thermal photons coincide.  So one can conclude, that both in 1+2
and 1+1 cases the system on initial stages cools predominantly due to
longitudinal expansion. For lower initial temperatures $T_{in}=500\, MeV$ and
especially $T_{in}=300\, MeV$ thermal photon spectra evaluated within 2+1
hydrodynamics are harder. This is evidence, that emission from later stages
with lower temperature, but higher radial velocity and thus high effective
temperature (\ref{teff}) dominate over emission from QGP.

Sensitivity of the yields to the variations of transition and freeze-out
temperatures are shown on the fig.\ \ref{2dbjor-ti}, right plot. Evaluations
with basic set of parameters (\ref{set-2dim}) (solid lines), higher
transition temperature $T_c=200\, MeV$ (dotted lines), and lower freeze-out
temperature $T_f=120\, MeV$ (dashed lines) are shown. In all three cases
decay photons goes very close to each other. The same are thermal
photons, however decreasing of $T_f$ slightly increase yield of intermediate
thermal photons.

In all previous calculations we used the simplest equation of state (EoS) --
EoS with first order phase transition with ideal gas of quarks and gluons
(degeneracy $g_{QGP}=37$) in one phase and ideal gas of massless pions
(degeneracy $g_{hadr}=3$) in the other phase. Such EoS includes large jump in
of specific entropy, and thus result in long living mixed phase. Besides this,
such EoS leads to maximal possible velocity of sound $v_c=\sqrt{1/3}$ both
in QGP phase and hadronic phase and thus maximal possible radial velocity at
fixed value and spatial distribution of initial energy. If one considers
massive hadrons instead of massless, and interacting quarks and gluons
instead of free ones, then one finds $v_c<\sqrt{1/3}$. In addition, this
inclusion of heavier hadrons and resonances in hadronic gas significantly
increase its degeneracy and reduce jump of entropy. These effects result in
decreasing of the yield of thermal photons in intermediate region.

\begin{figure}[ht]
\vspace*{9.1cm}
\includegraphics{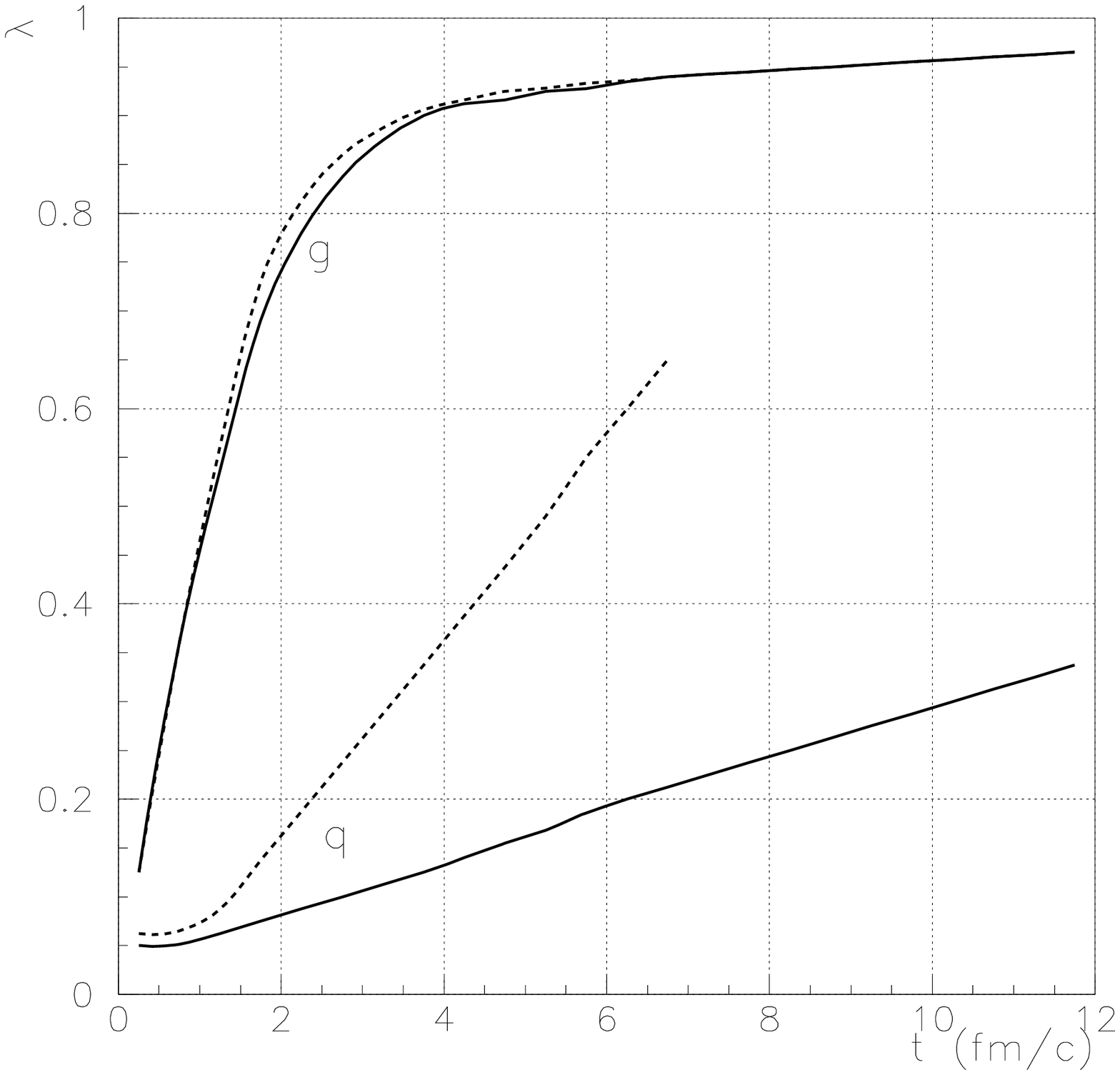}
\includegraphics{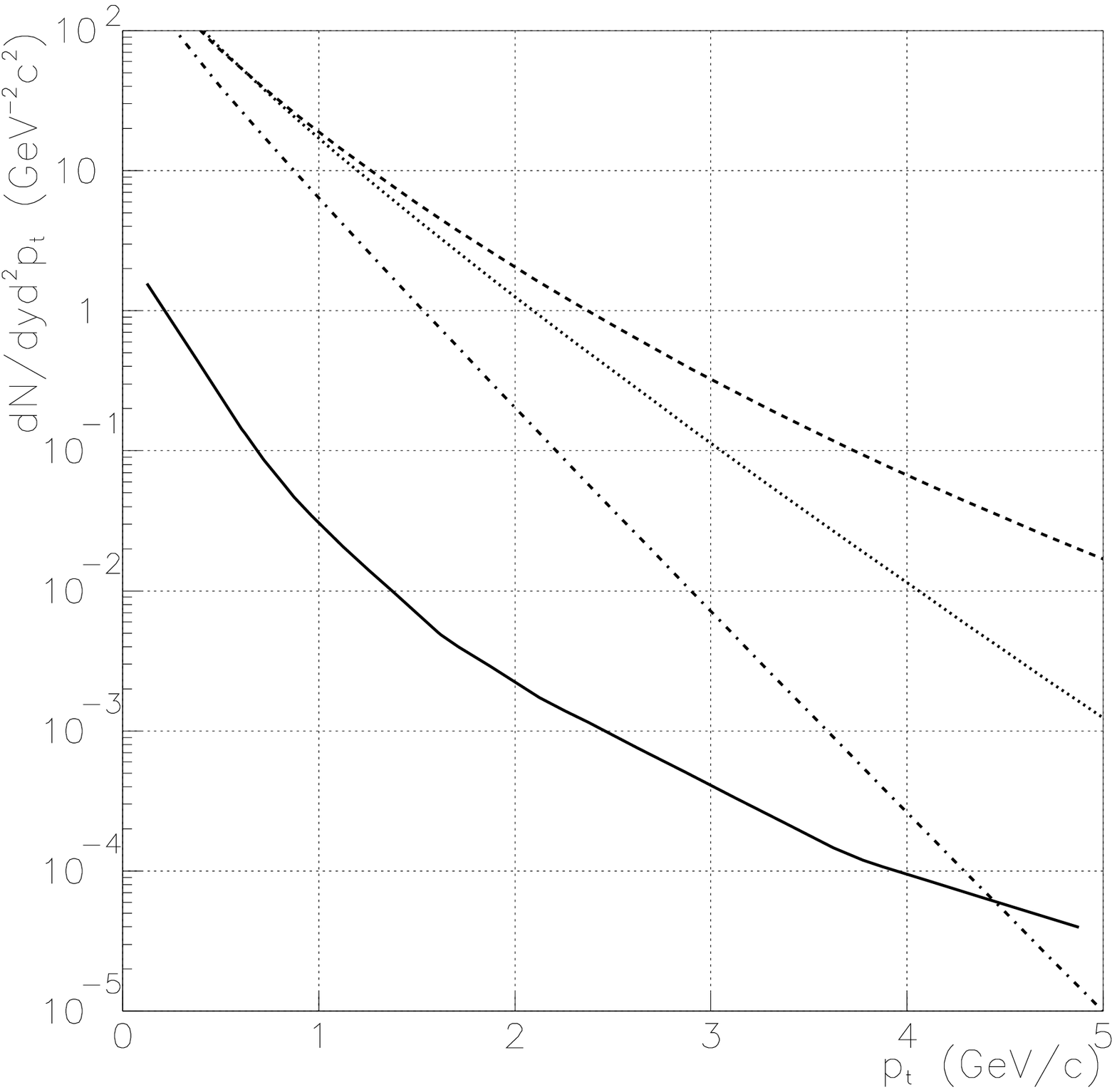}
\caption{Left plot: fugacity of quarks and gluons at LHC, solid lines --
         fixed $\alpha_s$, dotted lines -- running $\alpha_s$. Right plot:
         thermal photon yield from QGP without chemical equilibrium (solid
         line) and from 2+1 Bjorken hydrodynamics with $T_{in}=1\, GeV$
        (dashed line) $T_{in}=500\, MeV$ (dotted line) and $T_{in}=300\, MeV$
        (dash-dotted line). Only Compton scattering and annihilation
        (emission rate (\protect\ref{kapli})) are taken into account.}
\label{chem-eq-reach}
\end{figure}

Another phenomenon perhaps important in evaluation of direct photon emission
in ultrarelativistic heavy ion collision is absence of chemical equilibrium
in the QGP phase. It was argued \cite{rate-chem1,chem1,chem2,chem3} that in
the very beginning of collision quarks and gluons are strongly undersaturated
(approximately 0.1 of equilibrium multiplicity), and subsequently their
number is increased due to reactions $gg\leftrightarrow ggg$,
$gg\leftrightarrow q\bar q$ etc.  Because of comparatively small coupling
constant, chemical equilibration is reached rather slowly -- see fig.\
\ref{chem-eq-reach}, left plot, with results from ref.
\cite{chem3}. In these evaluations initial conditions are taken from HIJING
event generator, and then Boltzmann equation was solved with collisional
term including all possible QCD binary interactions as well as
reaction $gg\leftrightarrow ggg$, included to describe establishing of
chemical equilibrium. Solid lines on the fig.\ \ref{chem-eq-reach} correspond
to the evaluations with fixed $\alpha_s=0.3$, and dashed lines -- to
evaluations with running coupling constant

$$
\alpha_s(<\!E\!>)=\frac{4\pi}{9\ln\left(<\!E\!>^2/\Lambda^2_{QCD}\right)},
$$
where $<\!E\!>$ -- average energy of partons evolved. As one can see from
this figure, due to higher coupling, gluons come to the equilibrium faster
than quarks and rate of equilibration practically independent on $\alpha_s$.
In contrast to them quarks equilibrate much slower, and this process strongly
depends on $\alpha_s$. However, one should take these evaluations with care:
cross sections, entering collisional term has Coulomb divergency. To screen
them one has to introduce cuts on soft quark and gluon exchange from some
considerations, and result should be very sensitive to the chosen values of
these cuts.

Yield of thermal photons emitted from QGP, evaluated within this approach
with accounting of Compton scattering and annihilation and with time
dependent $\alpha_s$, is shown on fig.\ \ref{chem-eq-reach}, right plot, by
solid line. We compare this yield with one, evaluated within 2+1 Bjorken
hydrodynamics with different initial temperatures and using emission rate
(\ref{kapli}) i.e.  also including only Compton scattering and annihilation.
We find that in the case of absence of chemical equilibrium yield of thermal
photons decrease by $10^{-3}-10^{-1}$.

However absence of chemical equilibrium does not influence significantly on
the yield of thermal photons. As we have seen on fig.\ \ref{times}, due to
significant radial velocity contribution of the hadronic gas, it is
approximately the same as contribution of the hot phase. So even disappearing
of the emission from the hottest stage does not reduce total yield of thermal
photons more than several times.

%===========================================================================

\subsection{Thermal photons from QGP and hadronic gas}

From the point of view of registration of QGP it is important
to know, whether or not thermal photons from QGP dominate over the thermal
photons from hadronic gas. From one hand, thermal photons emitted from QGP
have highest temperature, from the other hand thermal photons from hadronic
gas have higher radial collective velocity, so that the answer is not
obvious. In the paper \cite{prompt-old} it was shown, that within 2+1 Bjorken
hydrodynamics and emission rate (\ref{kapli}) the contribution from QGP is
slightly smaller than contribution from hadronic gas. However, in the recent
paper \cite{lhc-high-rate}, emission of thermal photons was evaluated with a
rate accounting bremsstrahlung (\ref{twoloop-brems}) and annihilation with
scattering on third particle (\ref{twoloop-ann}). It was shown, that in this
case even for initial temperatures $T_{in}\sim 500\, MeV$ emission from QGP
dominate over emission from hadronic gas. It is interesting, that in this
case energy carried out by thermal photons becomes comparable with energy
carried by final hadrons, and evolution of the system should by corrected to
take into account energy loss due to thermal photon emission.

One can estimate yield of thermal photons from QGP and hadronic gas from
fig.\ \ref{times}. Third curve from the bottom corresponds to the time
($\tau < 10 fm$) of life of QGP phase. This curve should be compared with
total photon yield. In the case of initial temperature $T_{in}=1\, GeV$ and
$dN/dy=12000$ (left plot) contribution of QGP is approximately twice smaller
than total photon yield, what means, that contributions of QGP and hadronic
gas are comparable even at $p_t\sim 5\, GeV/c$. In the case $dN/dy=6000$
(right plot) relative contribution of hadronic gas is smaller and QGP photons
could dominate.

%=============================================================================

\subsection{Direct photons at SpS}

Below we compare spectra of thermal photons in Pb+Pb collisions at $158\,
A\cdot GeV$ (SpS), predicted by various models, with preliminary
experimental results, obtained by WA98 collaboration. Our goal is not to
choose model, which better describes experimental results or even to conclude,
whether or not QGP was formed in these collisions. We just would like to
demonstrate, that calculations within 2+1 Bjorken hydrodynamics can reproduce
measured spectrum of thermal photons, providing we guessed appropriate
initial conditions. As well we show, that at least at this energy different
models give very similar predictions, so it is possible, that at LHC energy
these predictions will be close too.

\begin{figure}[ht]
\vspace*{9.5cm}
\includegraphics{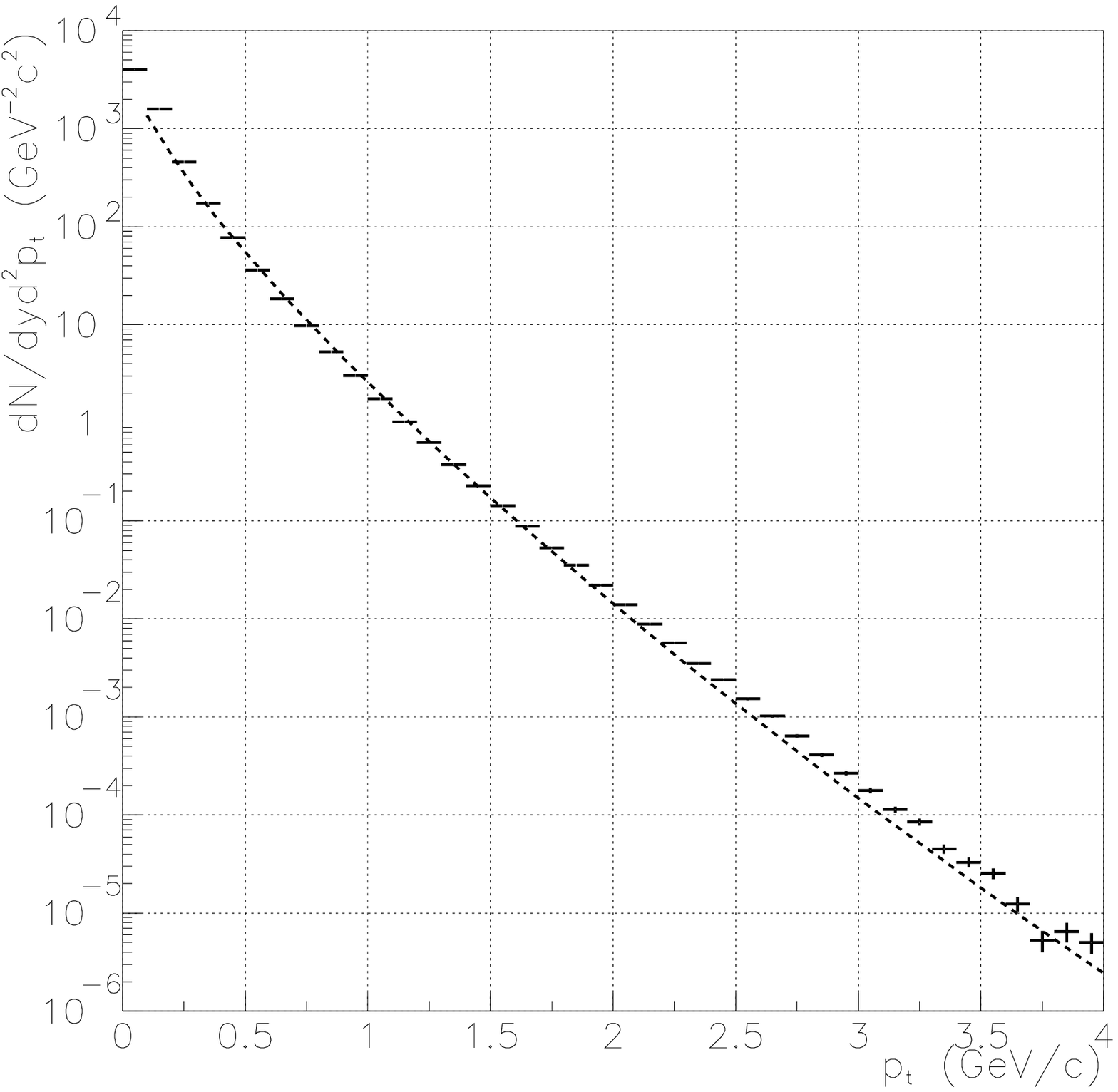}
\includegraphics{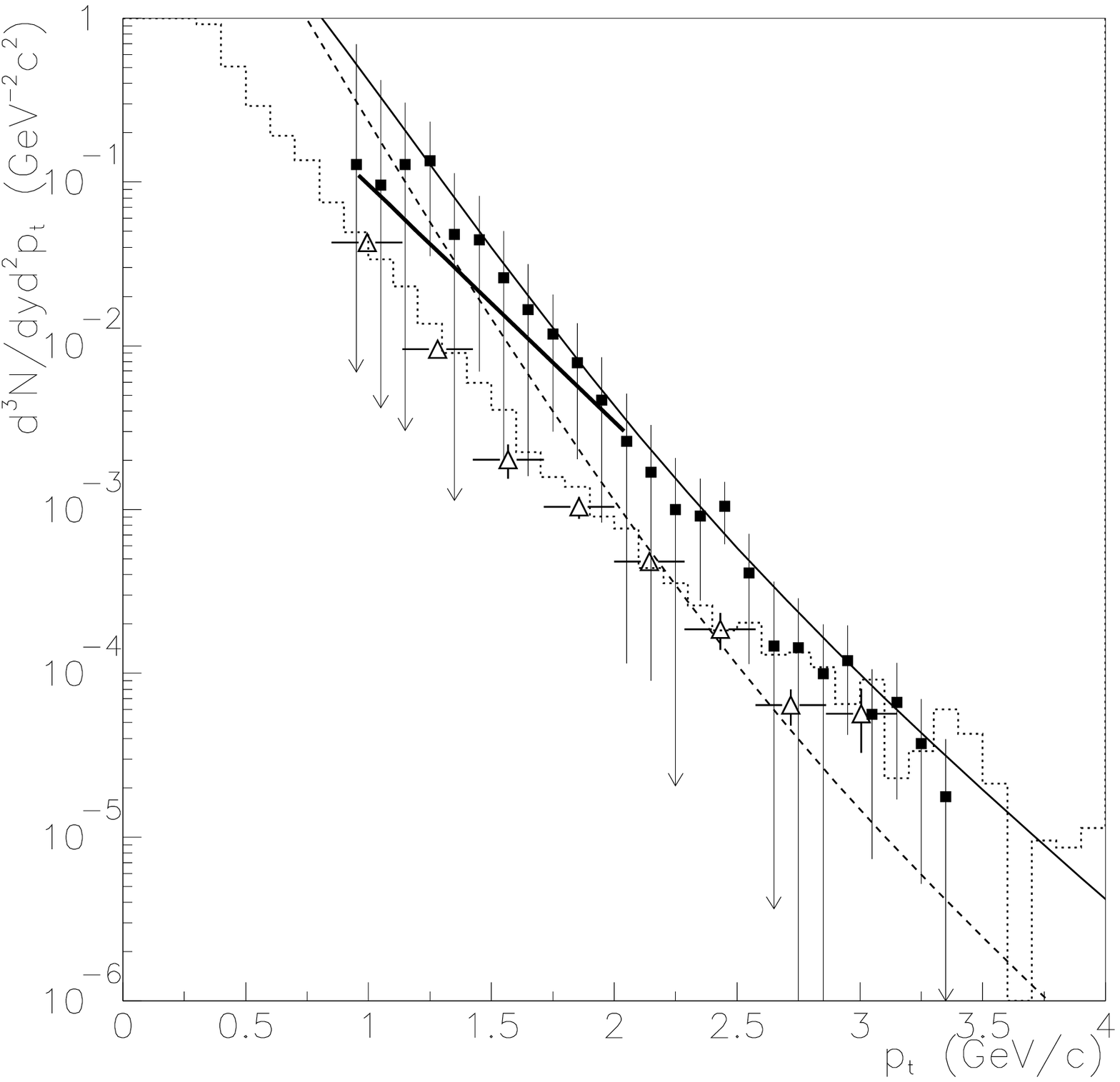}
\caption{Left plot: total photon yield in Pb+Pb collision at $158\, A\cdot
         GeV$, measured by WA98 collaboration, preliminary data, (crosses)
         and evaluated within 2+1 bjorken hydrodynamics (dashed line). Right
         plot -- yield of direct photons in this collision. Rectangles --
         preliminary WA98 results, dashed histogram -- predictions of Parton
         Cascade  Model, triangles -- predictions of UrQMD, thick solid line
         --  3+1 hydrodynamics, thin solid and dashed lines -- 2+1 Bjorken
         hydrodynamics with emission rate accounting of bremsstrahlung and
         contribution of $a_1$ resonance and without accounting of these
         effects correspondingly.}
\label{fig-WA98}
\end{figure}

Let us begin from 2+1 Bjorken hydrodynamics. In this case we fix $T_c=160\,
MeV$, $T_f=140\, MeV$ and choose initial conditions $T_{in}=320\, MeV$,
$\tau_{in}=0.4\,fm/c$ so that reproduce multiplicity $dN/dy=620$ and spectrum
\cite{WA98-tot}
of final photons. Quality of the data description can be seen on the fig.\
\ref{fig-WA98}, left plot. Having parameters of the model fixed, we evaluated
yield of thermal photons using two emission rates of thermal photons: in the
first case we use emission rate (\ref{kapli}) both in QGP and hadronic gas
phase and in the second case - emission rate (\ref{twoloop-brems}),
(\ref{twoloop-ann}) for QGP phase and (\ref{rate-a1-par}) for hadronic gas
phase. Evaluated yields of thermal photons are shown on the fig.\
\ref{fig-WA98}, right plot, by thin dashed line and thin solid line for the two
cases respectively. Experimental data are shown by rectangles.

Second hydrodynamic model we consider is fully hydrodynamic (3+1 dimensional)
model \cite{WA98-3dim}. Yield of thermal photons, evaluated within this model
is shown by thick solid line. In these evaluations emission rate
(\ref{kapli}) for QGP and (\ref{rate-a1-par}) for hadronic gas was used.
However, the difference between predictions of these models arises not from
type of longitudinal expansion -- scaling or hydrodynamic, but from initial
conditions: in the 3+1 hydrodynamics they are evaluated in spirit of
Landau hydrodynamics (shock waves propagate through the colliding nuclei and
heat up the system). Such description of initial stage of collision results
in higher initial temperatures compared to usually considered in Bjorken
scenarios, so one finds higher effective temperatures of thermal photons.

To compare predictions of equilibrium (hydrodynamic) models with
nonequilibrium (cascade) models we consider as well predictions of Partonic
Cascade Model (PCM) \cite{WA98-PCM} and Ultrarelativistic Quantum Molecular
Dynamic Model (UrQMD) model \cite{WA98-UrQMD}. PCM describes nucleus-nucleus
collisions in terms of parton picture of hadronic interactions. It performs
evolution using quantum kinetics with interactions of quarks and gluons,
evaluated within pQCD. Developed partonic picture clasterizes to form final
hadrons. Predictions of this model for $Pb+Pb$ collisions at $160\,A\cdot
GeV$, taken from \cite{WA98-PCM}, are shown on fig.\ \ref{fig-WA98}, right
plot, by dotted histogram. In contrast to PCM, the second cascade model,
UrQMD, deals with propagation and interactions of hadrons in hot matter.
Yield of direct photons in $Pb+Pb$ collisions at $160\,A\cdot GeV$ evaluated
within this model, taken from \cite{WA98-UrQMD}, is shown on
fig.\ \ref{fig-WA98}, right plot, by triangles.

Comparing predictions of all considered models one finds that they go rather
close to each other. Moreover, comparing PCM and UrQMD one finds a striking
coincidence of their predictions. It looks like the yield of direct photons
is independent on the details of the interactions of particles in the hot
matter. As for 2+1 Bjorken hydrodynamics, one can see, that yield, evaluated
with accounting of bremsstrahlung and annihilation with rescattering on third
particle gives the best description of experimental data among considered
models. So we conclude, that 2+1 Bjorken model provides reasonable
description of evolution of hot matter, such as we obtain good reproduction
of the yield of thermal photons.

Another important observation, we can made from fig.\ \ref{fig-WA98}, is that
in the range $p_t<5\, GeV/c$ predictions of all considered models differ less
than order of magnitude.

\section{Decay photons}

In the previous sections we evaluated yield of thermal photons. However from
the experimental point of view not only its absolute value but also ratio of
spectra of thermal and total photons is important. It is this ratio, which
controls possibility of extraction of thermal photons from the total photon
yield. To estimate total photon yield we have to evaluate spectrum of decay
photons, i.e. to evaluate $p_t$ distributions of $\pi^0$, $\eta$, etc.
hadrons, having channels of decay onto photons, and decay them into photons.
To do this we use several models: 2+1 Bjorken hydrodynamic model, HIJING~1.36
and VENUS~4.12 event generators.

\begin{figure}[ht]
\vspace*{9cm}
\includegraphics{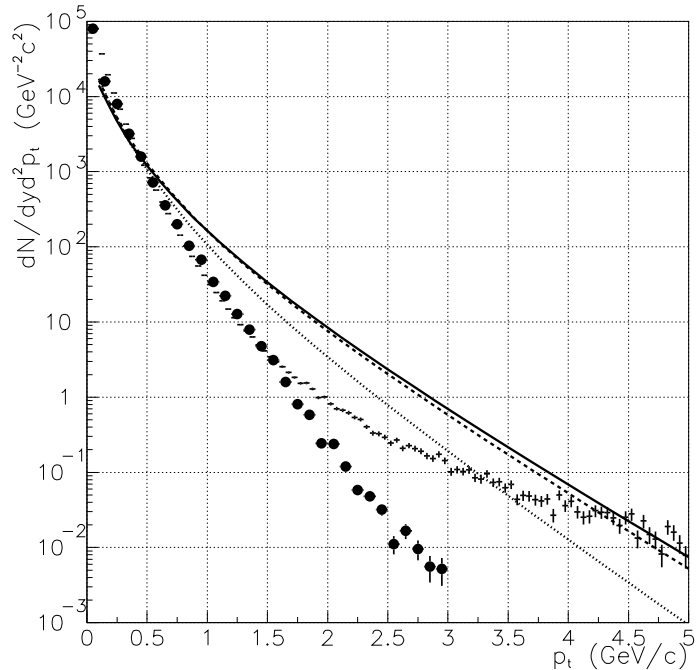}
\caption{Total photon yield in $Pb+Pb$ collision at LHC energy,
         evaluated within 2+1 Bjorken hydrodynamics with initial temperatures
         $T_{in}=1\, GeV$ (solid line), $T_{in}=500\, MeV$ (dashed line) and
         $T_{in}=300\, MeV$ (dotted line). Crosses -- predictions of
         HIJING~1.36, circles -- predictions of VENUS~4.12. }
\label{fig-back}
\end{figure}

On the fig.\ \ref{fig-back} we present results of hydrodynamic calculations
of yields of decay photons for three values of initial temperature
$T_{in}=1\, GeV$, $T_{in}=500\, MeV$ and $T_{in}=300\, MeV$, and compare them
with predictions of HIJING~1.36 and VENUS~4.12 models.

HIJING is model of heavy ion collisions, which consider initial stage of
collision (jet and minijet production) on the basis of perturbative QCD, and
describe fragmentation of created partons using Lund string model
\cite{HIJING}. HIJING does not include interactions of the secondary
particles. This results in nonexponential $p_t$ distribution with excess of
high energy and lack of intermediate energy particles in comparison with
thermal distribution. In contrast to HIJING, VENUS takes into account
rescattering of final particles \cite{VENUS} and thus results in exponential
spectrum, closer to the predictions of hydrodynamic model. However, VENUS
does no include radial flow, so that spectrum predicted by this model is softer
than one, predicted by hydrodynamics. In evaluations of the ratio
thermal to decay photons we use predictions of spectrum of decay
photons from hydrodynamic model and HIJING.

\begin{figure}[ht]
\vspace*{9.5cm}
\includegraphics{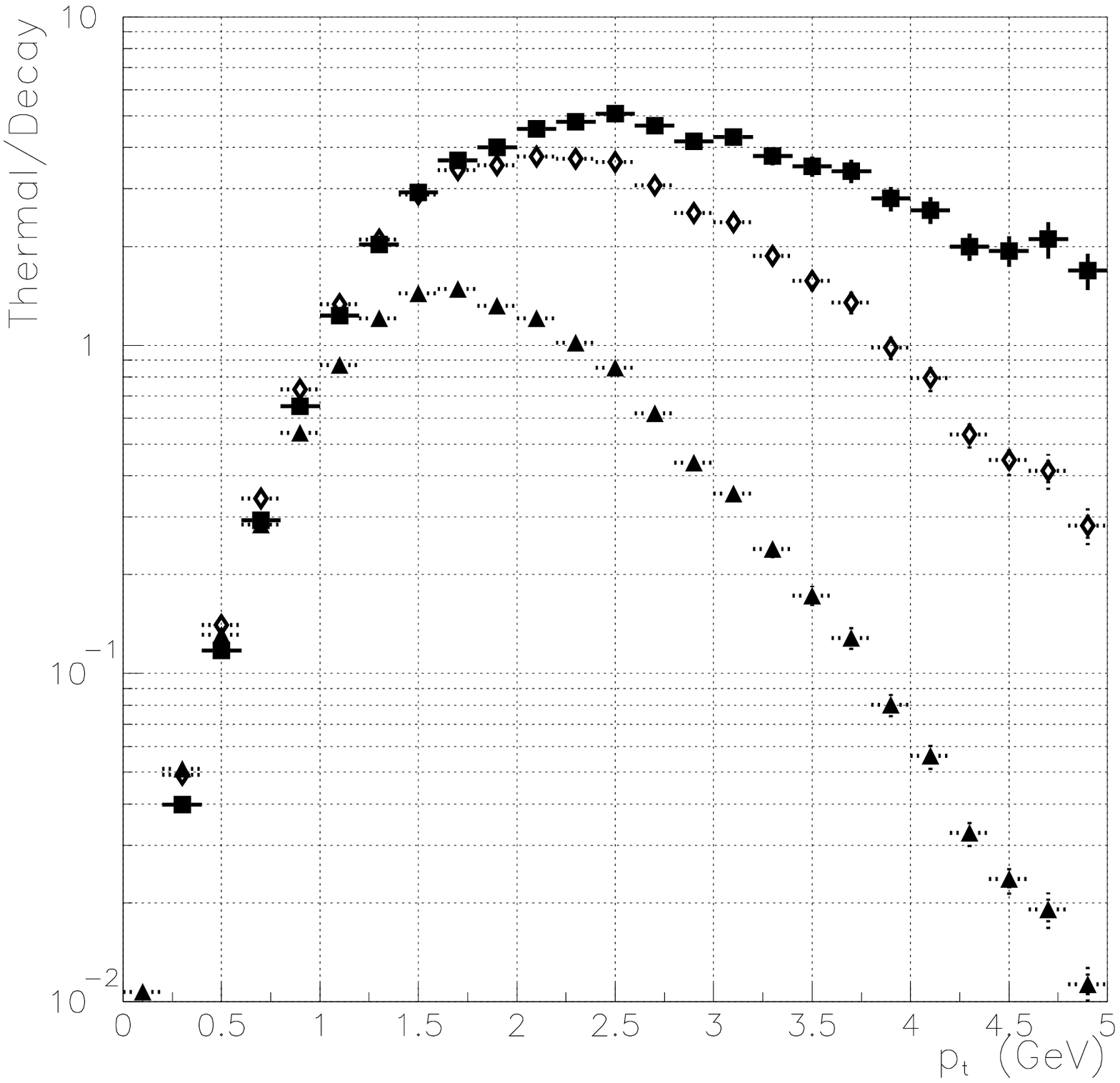}
\includegraphics{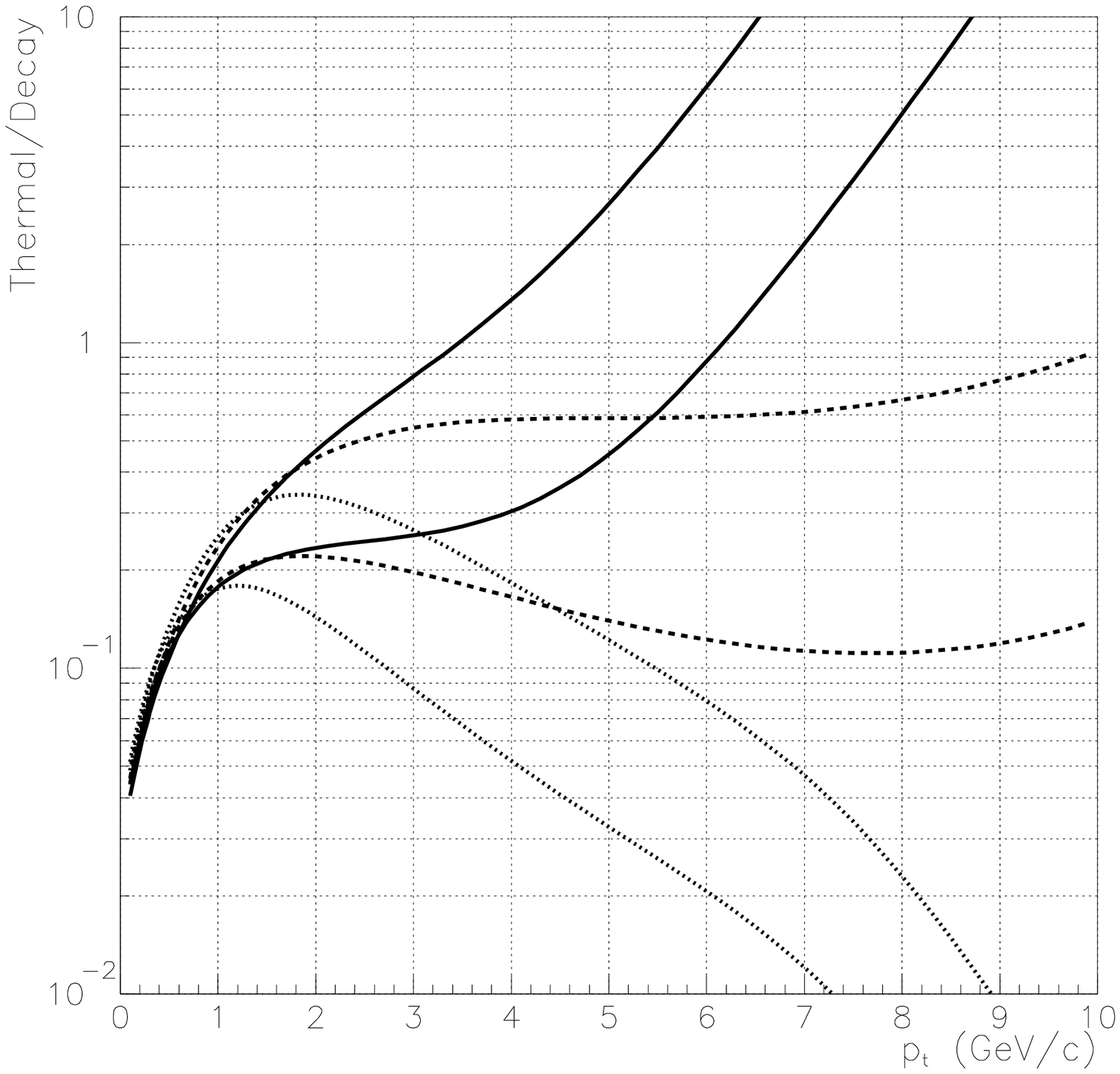}
\caption{Ratio yield of thermal to yield of decay photons in $Pb+Pb$
         collision at LHC energy. Left plot: Thermal photons, evaluated
         within 2+1 Bjorken hydrodynamics, are divided by decay photons,
         taken from HIJING. Rectangles correspond to thermal photons,
         evaluated with initial temperatures $T_{in}=1\, GeV$, diamonds --
         $T_{in}=500\, MeV$ and triangles -- $T_{in}=300\, MeV$, in
         all cases emission rate (\protect\ref{kapli}) was used. Right
         plot: both thermal and decay photons are evaluated within 2+1
         Bjorken hydrodynamics. Solid lines correspond to initial
         temperature $T_{in}=1\, GeV$, dashed -- $T_{in}=500\, MeV$ and
         dotted -- $T_{in}=300\, MeV$, lower lines evaluated with emission
         rate (\protect\ref{kapli}), upper -- with emission rates
         (\protect\ref{twoloop-brems}), (\protect\ref{twoloop-ann}) and
         (\protect\ref{rate-a1-par}).  }
\label{fig-ratio}
\end{figure}

If one uses $p_t$ distribution of decay photons, given by HIJING~1.36,
(see fig.\ \ref{fig-ratio}, left plot) then one finds, that thermal photons
dominate over decay photons in the intermediate region for all considered
initial temperatures. This takes place because of rather special shape of
$p_t$ distribution of final hadrons, predicted by HIJING. In contrast to
this, if one evaluates both thermal and decay photons within
hydrodynamic model (see fig.\ \ref{fig-ratio}, right plot), then for
initial temperatures $\sim 300\, MeV$ there will be region, where ratio
reached its maximal value and position of the maximum is shifted to hard part
with increasing of initial temperature, while at $T_{in}\sim 500\, MeV$
becomes approximately constant. For higher initial temperatures contribution
of thermal photons increase with increasing of $p_t$.

If one consider only leading order $\alpha_s$ reactions in QGP -- Compton
scattering and annihilation, then the ratio Thermal/Total reaches 20\% in
maximum even for lower considered initial temperature. If one includes higher
order $\alpha_s$ corrections -- bremsstrahlung ann annihilation with
rescattering on third particle, then the ratio Thermal/Total, evaluated for
$T_{in}=300\, MeV$ reaches 30-40\%.

\section{Conclusion}

We have considered emission of prompt, thermal and decay photons in
$Pb+Pb$ collisions at SpS and LHC energies and estimated uncertainty in
predictions of their yields. We find, that main uncertainty in the yield of
prompt photons comes from the lack of precise knowledge of structure
functions at small $x$ and their modifications in nuclei. This leads to
variations of the yield of prompt photons within factor $\sim 3$.

For thermal photons situation is more complicated. The main source of
uncertainty comes from impossibility to describe uniquely evolution of heavy
ion collisions. For description of evolution the both kinetic and hydrodynamic
models are used. We compared predictions of kinetic and hydrodynamic models
at SpS energy, however for LHC energy we concentrated on hydrodynamic models.
We find, that hard part of the spectrum $p_t>4-5 \,GeV$ slightly depends on
the presence of the radial flow (compare figs.\ \ref{1dbjor-np} for 1+1
expansion and \ref{2dbjor-ti} for 2+1). This takes place because this region
of $p_t$ is populated due to emission from the initial stage (see fig.\
ref{times}) when radial flow is not developed yet. Even if extremely high
initial temperatures are created in the beginning of the collision, they are
slightly affect on the radial flow, because as a result of longitudinal
expansion the high temperatures lives very short time and can not accelerate
considerably the matter in radial direction. This is illustrated on the fig.\
\ref{2dbjor-ti}:  difference of the effective temperatures of decay
photons, evaluated at $T_{in}=1\, GeV$ and $T_{in}=500\, MeV$ is smaller,
than difference between $T_{in}=500\, MeV$ and $T_{in}=300\, MeV$. So, we
find that for hard part of spectrum 1+1 and 2+1 hydrodynamics models give
similar predictions. However, 2+1 Bjorken hydrodynamics is not quite
two-dimensional -- longitudinal expansion in it is scaling. Comparing
predictions of 1+1 Bjorken and 1+1 Landau hydrodynamics we show, that this is
not important for the case of thermal photons.

In addition to ambiguity in description of evolution, there is significant
uncertainty in the emission rate: inclusion of the higher order
$\alpha_s$ corrections result in increasing of it up to order of magnitude.

We show, that hard part of thermal photon distribution is sensitive to the
scenario of collision. If fast thermalization of colliding nuclei with
high initial temperature takes place, than one finds high yield of thermal
photons, otherwise, if string-tube model of thermalization takes place, then
yield of thermal photons in hard part will be much smaller.

We estimated the ratio of Direct/Decay photons at LHC energy and find it
large enough ($20-30\%$) to be measured by ALICE PHOS setup, whose expected
sensitivity to the direct photons is approximately 5\% \cite{PHOS}. So that
thermal photons considerably contribute into total photon yield. If thermal
photons from QGP are evaluated without accounting of bremsstrahlung and
annihilation with scattering on third particle, then yield of thermal photons
from hadronic gas exceeds yield of thermal photons from QGP
\cite{prompt-old}. Otherwise, if one takes into account these higher order
$\alpha_s$ corrections, then thermal photons from QGP dominates over photons
from hadronic gas \cite{lhc-high-rate}.

The authors would like to thank V.I. Manko for interest and useful
discussions. This work was supported by grant RFFI 96-15-96548.

\end{document}